\documentclass{aa}  

\usepackage{graphicx}
\usepackage{txfonts}
\usepackage{amsfonts,amsmath,amssymb}
\usepackage{array,booktabs}
\usepackage{natbib}
\usepackage{color}
\usepackage{ulem}
\definecolor{teal}{RGB}{0,128,128}
\definecolor{mygray}{gray}{0.6}

\begin{document} 
   
\title{Flux dependence of redshift distribution and clustering of LOFAR radio sources}

\author{Nitesh~Bhardwaj
        \inst{1}\thanks{nitesh@physik.uni-bielefeld.de}
        \and
        Dominik~J.~Schwarz\inst{1}
        \and
        Catherine~L.~Hale\inst{2,3} 
        \and
        Kenneth~J.~Duncan\inst{2}
        \and 
        Stefano~Camera\inst{4,5,6,7}
        \and
        Caroline~S.~Heneka \inst{8}
        \and
        Szymon~J.~Nakoneczny\inst{9,10}
        \and
        Huub~J.~A.~R\"ottgering\inst{11}
        \and
        Thilo~M.~Siewert\inst{1}
        \and
        Prabhakar~Tiwari\inst{12}
        \and 
        Jinglan~Zheng\inst{1}
        \and
        George~Miley\inst{11}
        \and 
        Cyril~Tasse\inst{13,14} 
        }

    \institute{Fakult\"at f\"ur Physik, Universit\"at Bielefeld, Postfach 100131, 33501 Bielefeld, Germany
    \and
    School of Physics and Astronomy, Institute for Astronomy, University of Edinburgh, Royal Observatory, Blackford Hill, EH9 3HJ Edinburgh, UK
    \and
    Astrophysics, University of Oxford, Denys Wilkinson Building, Keble Road, Oxford, OX1 3RH, UK
    \and
    Dipartimento di Fisica, Università degli Studi di Torino, via P. Giuria 1, 10125 Torino, Italy
    \and
    INFN -- Istituto Nazionale di Fisica Nucleare, Sezione di Torino, Via P.\ Giuria 1, 10125 Torino, Italy
    \and
    INAF -- Istituto Nazionale di Astrofisica, Osservatorio Astrofisico di Torino, Strada Osservatorio 20, 10025 Pino Torinese, Italy
    \and
    Department of Physics \& Astronomy, University of the Western Cape, Cape Town 7535, South Africa
    \and 
    Institute for Theoretical Physics, Heidelberg University, Philosophenweg 16, 69120 Heidelberg, Germany
    \and
    Division of Physics, Mathematics and Astronomy, California Institute of Technology, 1200 E California Blvd, Pasadena, CA 91125, USA 
    \and 
    Department of Astrophysics, National Centre for Nuclear Research, Pasteura 7, 02-093 Warsaw, Poland
    \and
    Leiden Observatory, Leiden University, PO Box 9513, 2300 RA Leiden, The Netherlands
    \and
    Department of Physics, Guangdong Technion - Israel Institute of Technology, Shantou, Guangdong 515063, P.R. China
    \and
    GEPI \& ORN, Observatoire de Paris, Université PSL, CNRS, 5 Place Jules Janssen, 92190 Meudon, France 
    \and 
    Department of Physics \& Electronics, Rhodes University, PO Box 94, Grahamstown, 6140, South Africa
    }

\date{Received month date, year; accepted month date, year}

\abstract
{} 
{We study the flux density dependence of the redshift-distribution of low-frequency radio sources observed in the LOFAR Two-metre Sky Survey (LoTSS) deep fields and apply it to estimate the clustering length of the large-scale structure of the Universe, examining flux density limited samples (1~mJy, 2~mJy, 4~mJy and 8~mJy) of LoTSS-DR1 radio sources.}
{We utilize and combine the posterior probability distributions of photometric redshift determinations for LoTSS Deep Field observations from three different fields (Boötes, Lockman Hole and ELAIS-N1, together about $26$ square degrees of sky), which are available for between $91\%$ to $96\%$ of all sources above the studied flux density thresholds and observed in the area covered by multi-frequency data. We estimate uncertainties by a bootstrap method. We apply the inferred redshift distribution on the LoTSS wide area radio sources from the HETDEX field (LoTSS-DR1; about $424$ square degrees) and make use of the Limber approximation and a power-law model of three dimensional clustering to measure the clustering length, $r_0$, for various models of the evolution of clustering.}
{We find that the redshift distributions from all three LoTSS deep fields agree within expected uncertainties. We show that the radio source population probed by LoTSS at flux densities above $1$ mJy has a median redshift of at least $0.9$. At $2$ mJy, we measure the clustering length of LoTSS radio sources to be $r_0 = (10.1\pm 2.6) \ h^{-1}$Mpc in the context of the comoving clustering model.}
{Our findings are in agreement with measurements at higher flux density thresholds at the same frequency and with measurements at higher frequencies in the context of the comoving clustering model. 
Based on the inferred flux density limited redshift distribution of LoTSS deep field radio sources, the full wide area LoTSS will eventually cover an effective (source weighted) comoving volume of about $10\,  h^{-3}$ Gpc$^3$.}


\maketitle


\section{Introduction}

The excellent multi-frequency coverage of the LOFAR Two-metre Sky Survey (LoTSS) Deep Fields provides an opportunity to learn about the redshift distribution of low-frequency radio sources (120 MHz to 168 MHz). In turn the redshift distribution of radio sources is an essential ingredient in the study of the spatial clustering of radio sources and their evolution. The LoTSS Deep Fields first data release \citep[LoTSS-DF-DR1;][]{Tasse2021,Sabater2021} offers information such as the type of sources \citep{Best_2023}, 
cross-matching with multi-frequency observations \citep{Kondapally2021} and an improved approach on estimating the probability distribution functions (pdfs) of photometric redshifts (photo-z) of radio sources \citep{Ken21}, resulting in a high level of photo-z completeness. Photo-z information derived from the full posterior pdf and 
spectroscopic redshift information if available have been included in released catalogues\footnote{http://cdsarc.u-strasbg.fr/viz-bin/cat/J/A+A/648/A4}; the full posterior redshift distributions for the individual sources are not publicly available.

Continuum radio surveys enable us to study the angular distribution of the large-scale structure of the Universe, see \citet{Longair, Becker95, Condon98, Miley} for early works.
Obtaining redshift information is essential for the study of the corresponding spatial large-scale distribution and to extract cosmological parameters \citep{Camera2012}, especially as radio galaxies exist over a wide range in redshift \citep{drinkwater_schmidt_1996, refId0, Best_2023}. 
Thus, one needs to consider large radio surveys and complement them with redshift measurements for -- if possible -- all radio sources. This is a challenging 
task and one is limited by the number of sources in such surveys which have an estimate of their photometric or spectroscopic redshift. Usually, redshift estimates cannot be obtained for each source in wide area surveys, but are available for smaller deep fields with good multi-frequency or good spectroscopic coverage (see e.g.\ \citealt{refId0, Kondapally2021}).
While flux density-limited continuum radio surveys can provide angular positions for a wide area of sky, the full three dimensional analysis of the cosmic structure requires to measure at least the statistical distribution of radio sources as a function of redshift or distance above a given flux density, expressed by a pdf, $p(z)$, i.e.
\begin{equation}
\mathrm{d}n = \bar n \tilde{p}(r) \mathrm{d}r = \bar n p(z) \mathrm{d}z, 
\end{equation}
where $\bar n$ denotes the average surface density of a flux density limited survey, $r$ denotes the comoving radial distance and $z$ the cosmological redshift. The functions $\tilde{p}(r)$ and $p(z)$ denote the sample pdfs in 
comoving radial distance and redshift space, respectively. With an estimate of $p(z)$ in hand, we can then infer statistical properties of the three dimensional large scale structure from the projected two dimensional information contained in a wide area continuum radio survey.

Previous studies of the clustering properties and redshift distribution of radio sources were done primarily for radio continuum surveys at frequencies around $1$ GHz. The angular clustering properties of the the VLA Faint Images Of the Radio Sky at Twenty centimeters survey \citep[FIRST;][]{Becker95} and the 
NRAO VLA Sky Survey \citep[NVSS;][]{Condon98} have been studied extensively (see e.g.\ \citealt{Cress1996,BlakeWall2002,Overzier2003,NusserTiwari2015,ChenSchwarz16}).
However, those studies have been limited by quite restricted knowledge of the redshift distribution of radio selected samples, as e.g.\ the Combined EIS-NVSS Survey of Radio Sources \citep[CENSORS;][]{Brookes_2008} provided spectroscopic follow up of just 143 NVSS radio sources. Thus, extensive use has been made of optical galaxy redshift surveys, such as the 6 degree Field Galaxy Survey (6dFGS; \citealt{6dFGS}) in \citet{MauchSadler2007}, covering the redshift range $0.003 < z < 0.3$. Cross-matching of NVSS, FIRST and other radio surveys with catalogues from the Sloan Digital Sky Survey (SDSS; \citealt{Eisenstein2001}) produced matches for about a third of all radio objects, still for redshifts below one, see e.g.\ \citet{Kimball2008, Donoso2009}.
A better understanding of the redshift distribution of also high redshift radio sources required deep fields with good multi-wavelength coverage, 
like the COSMOS field \citep{2007ApJS..172....1S, refId0}.
At frequencies well below $1$ GHz and similar to the LoTSS frequency range, the TIFR-Two Meter Sky Survey Alternative Data Release~1 \citep[TGSS;][]{TGSS2017} provided the first opportunity for a wide survey area to estimate the redshift distribution of radio sources via a cross matching with SDSS quasars \citep{SDSSDR14} and to study the angular clustering properties at low radio frequencies \citep{Dolfi2019}. 
The $p(z)$ obtained in these studies are different from our measurements in the sense that we make use of 
photometric measurements for about 95\% of all observed radio sources.

In this work we study the distribution of low-frequency radio sources as a function of redshift and flux density by combining observations from three deep survey regions. We infer the redshift distribution of the radio source sample by combining their individual posterior pdfs, which were derived from multi-wavelength observations of the three deep fields in \citet{Ken21}\footnote{To be made available on Vizier after publication of this work.}. We then make use of this flux density limited redshift distribution and the corresponding angular two-point correlation measurements \citep{siewert} from the wide field LoTSS data release~1 radio sources in the HETDEX field \citep[LoTSS-DR1;][]{Shimwell19} to infer clustering properties such as the correlation length of these radio sources. 
We make use of the \textit{value added source} catalogue of LoTSS-DR1, in which artifacts and multiple components of radio sources have been identified \citep{Williams2019}. Note that we use deep and wide field data for which artifacts of bright sources have largely been removed by a major effort of visual inspection by experts, in contrast to the more recent second data release of LoTSS \citet{Shimwell2022}. The LoTSS-DR1 value added source catalogue contains photometric redshifts for about~$48$\% of all sources. Mainly the limited depth of the multi-wavelength data used in the LoTSS analysis in the HETDEX field \citep{Duncan2019} gives rise to selection effects that prevent us from the direct application of the resulting redshift distribution on all radio sources.

The clustering measurements from LoTSS-DR1 are presented in \citet{siewert}. 
After carefully accounting for survey masks, systematics, and artifacts arising from multiple components of radio sources, they achieve clustering measures for radio sources that reasonably align with standard $\Lambda$CDM cosmology. A complementary clustering study that is based on the wider LoTSS-DR2 \textit{radio source} catalogue \citep{Shimwell2022} is presented in \citet{Hale_2023}. In \citet{Nakoneczny_2023} the cross-correlation of LoTSS-DR2 radio sources with the cosmic microwave background is studied. The analysis of \citet{Hale_2023} and \citet{Nakoneczny_2023} fixes the 
flux density threshold to $1.5$ mJy and is based on the method to infer the overall redshift distribution of LoTSS Deep Field sources presented and discussed in detail in this work, but also differs in order to make use of both spectroscopic (for 26\% of all deep field sources) and photometric redshifts. In this work we decided not to make use of any spectroscopic information. The selection function for the available spectroscopic data of cross-matched radio sources in the three deep fields is unknown. Ideally they would be drawn from a random sample of radio sources. In order to avoid such biases we stick to the photometric data, which are sampled homogeneously in each of the three fields. We assume that the remaining systematic uncertainties of the photometry are captured by the differences in multi-wavelength coverage and data quality between the three fields, which we account for by bootstrap sampling as described below.

\begin{table*}
	\centering
    \caption{Number of sources ($N$) in different fields for different flux density thresholds ($S_\mathrm{min}$) and the fraction of sources ($f$) which have photometric redshift pdfs and spectroscopic redshift measurements (not used in this work, but included in \citealt{Hale_2023,Nakoneczny_2023}), respectively. \label{tab1}}
\begin{tabular}{cccccccccc}
\hline\hline

$S_\mathrm{min}$ & Boötes & Boötes & Boötes & LH & LH & LH & EN1 & EN1 & EN1 \\
 mJy & $N$ & $f_\mathrm{photo}$ & $f_\mathrm{spec}$ & $N$ & $f_\mathrm{photo}$ & $f_\mathrm{spec}$ & $N$  & $f_\mathrm{photo}$ & $f_\mathrm{spec}$ \\
\hline
0.0  & 19179 & 0.95 & 0.21 & 31162 & 0.97 & 0.05 & 31610 & 0.96 & 0.05 \\
0.5  & 7991 & 0.95 & 0.28 & 9356 & 0.96 & 0.10 & 5591 & 0.96 & 0.16 \\
1.0  &  2939 & 0.93 & 0.31 & 3464 & 0.94 & 0.17 & 2091 & 0.94 &  0.26\\
1.5  & 1848 & 0.92 &  0.31 & 2169 & 0.93 & 0.18 &  1287& 0.94 & 0.31 \\
2.0  & 1379 & 0.91 & 0.31 & 1617 & 0.92 & 0.19 & 968 & 0.93 & 0.34 \\
4.0  & 791 & 0.92 & 0.30& 948 & 0.91 &  0.19 & 555  & 0.93 & 0.36 \\
8.0  & 491 & 0.92 & 0.29& 615 & 0.89 & 0.20 & 370 & 0.93 & 0.41\\
\hline    			
\end{tabular}
\end{table*}

Throughout, we assume the spatially flat Lambda Cold Dark Matter (LCDM) model to convert the redshift of an object to a spatial distance (and vice versa) and use $\Omega_\mathrm{M} = 0.317$, in agreement with the Planck best-fit parameters \citep{2020A&A...641A...6P}. This work is structured as follows.
In the next section we describe the LoTSS-DF-DR1 data and obtain the redshift distribution for flux density limited samples from the measured posteriors of photometric redshifts, as presented in \citet{Ken21}. We describe our technique of weighted stacking of redshift pdfs of sources from the three aforementioned fields. 
In section 3 we summarise and extend some of the results on the angular two-point correlation function from \citet{siewert} with the wide field LoTSS-DR1 \citep{Shimwell19}. We estimate the clustering length in section 4. Finally, in section 5 we present our conclusions.

\section{Redshift distribution from LoTSS Deep Fields}

We consider the task of obtaining a redshift distribution function from the LoTSS-DF-DR1 \citep{Tasse2021,Sabater2021}.
Located in some of the best-studied northern extragalactic survey fields --- Boötes,
European Large Area Infrared Survey field North 1 (ELAIS-N1 or EN1), and the Lockman Hole (LH) ---  the LoTSS Deep Fields data reach a rms sensitivity of $\sim$ 32, 20, 22 $\mu$Jy/beam at a central frequency of 144 MHz for Boötes and LH, and at 146 MHz for EN1, respectively \citep{Tasse2021,Sabater2021}.

For the three deep fields multi-wavelength observations are available for different fractions of field of view. They cover the infrared, optical and X-ray and together allow us to identify and match 96\% of the radio sources within about 26 square degrees of sky \citep{Kondapally2021}. 
In all three fields, the multi-wavelength matched aperture photometry used for source identification and photometric redshift analysis spans from the UV to mid-infrared, however the exact set of filters and their associated sensitivity varies from field to field \citep[][see also Fig.~1 of \citeauthor{Ken21}~\citeyear{Ken21}]{Kondapally2021}.

\begin{figure}
     \centering
     \includegraphics[width= 0.5 \textwidth]{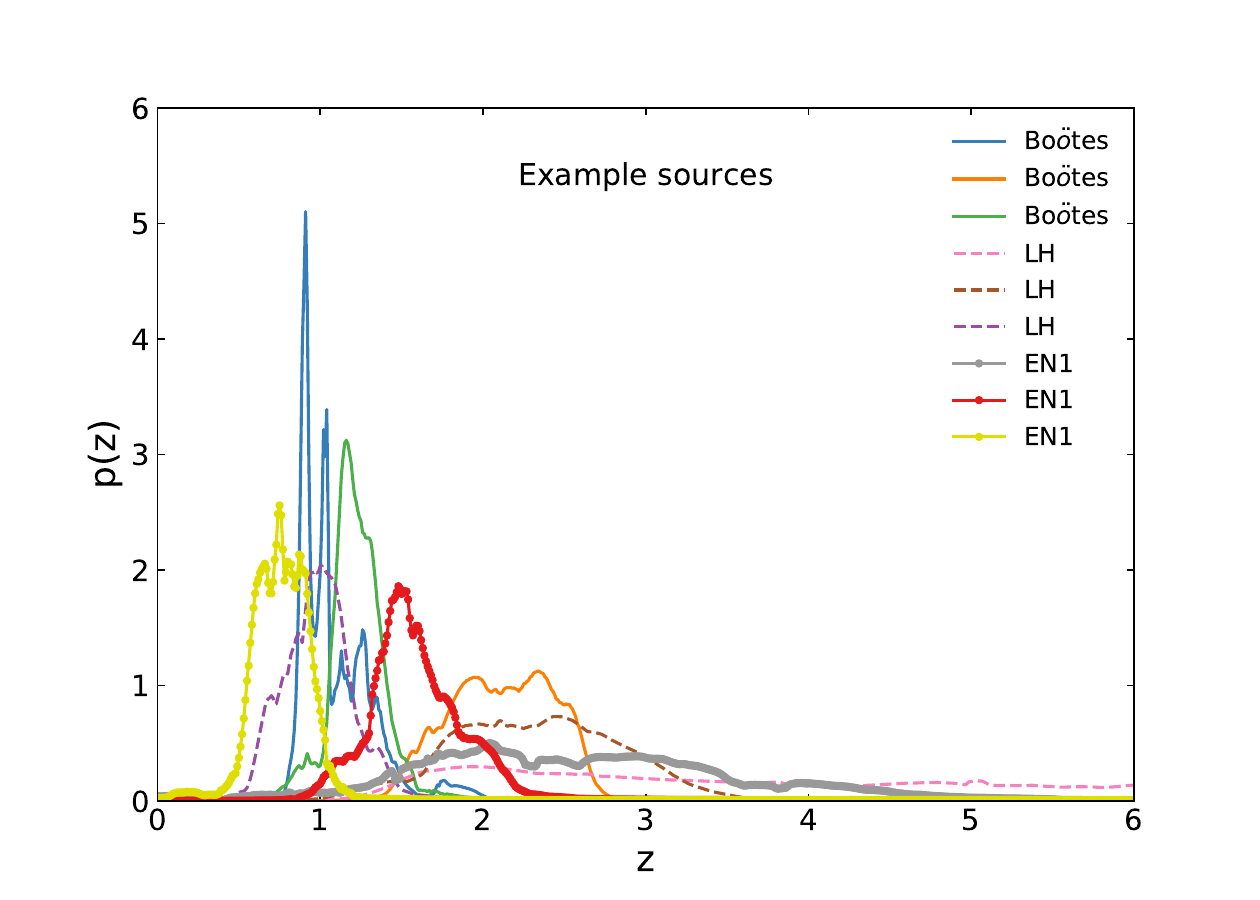}
     \caption{Example photo-z redshift posterior distributions of some sources from each of the three LoTSS Deep Fields (coded by colour and line-style).}
     \label{fig: pdf_single}
    \end{figure}

The photometric redshifts were obtained using a hybrid method that combines both template fitting and machine-learning estimates to produce a consensus redshift estimates and associated calibrated uncertainties. The full methodology is presented in \citet{Ken21}, here we briefly summarise the implementation. 
Three different template based estimations are calculated using the EAZY software \citep{Brammer2008} with three different template sets chosen to represent a range of different spectral energy distributions expected in the radio population, including both stellar only emission and combined stellar and active galactic nuclei (AGN) emission \citep{Duncan2018a}. The individual template fitting results are separately optimized using zero-point offsets calculated from the spectroscopic redshift sample in each field and the posterior redshift predictions calibrated such that they accurately represent the uncertainties in the estimates.
Next, additional machine-learning estimates are produced using the Gaussian process redshift code, GPz \citep{Almosallam2016a,Almosallam2016b}, with training and prediction performed separately for each field using the respective photometry and spectroscopic training samples.

Finally, the individual template and machine-learning estimates are then combined following the hierarchical Bayesian combination method presented in \citet{Dahlen2013}, incorporating the additional improvements outlined in \citet{Duncan2018a, Duncan2018b}.
The consensus photometric redshift posteriors for an individual galaxy, $p_{i}(z)$, are evaluated onto a grid based on the initial redshift steps used for template fitting, spanning from $0 \leq z \leq 7$. A sample of photometric redshift pdfs for nine randomly selected sources, three from each of the three deep fields, is shown in Fig.~\ref{fig: pdf_single}. As the figure demonstrates, the posterior pdf of many sources has a well defined peak, e.g.\ the sources indicated by the green full line, the red line with dots, or the purple dashed line, while other postreior pdfs are multimodal, e.g.\ the sources shown by the blue and orange full lines. For other sources, like the ones indicated by the pink dashed line and the grey line with dots, it is clear that they are at $z> 1$, with a broad redshift distribution.

Table \ref{tab1} shows the number of radio sources for various flux density thresholds per deep field and the fraction of sources with photometric and spectroscopic redshifts. Note that these numbers do not include any quality assessments, besides the mere existence of the posterior photo-z estimate. The degree of completeness of the photometric redshifts decreases with increasing flux density from 96\% below 0.5 mJy to 91\% at 8 mJy. The brighter sources are almost exclusively AGN \citep{Best_2023}, a population that extends to high redshifts where multiwavelength data become incomplete. Introducing a flux density threshold of at least 0.5 mJy, the fraction of spectroscopic redshifts lies between 10\% and 41\%, depending on the field and its flux density threshold. Note that while in the EN1 field the overall fraction of spectroscopic redshifts is as low as 5\% without a flux density thereshold (besides the source detection criterion of a signal to noise ratio of 5), for a flux density limit of 8\ mJy spectroscopic redshifts are available for 41\% of all radio sources. For the Lockman Hole we observe a similar trend, but reach only a 
completeness of 20\% at the highest flux density threshold. In contrast, also less complete at the highest flux densities, in the case of the Bo\"otes field, the fraction of radio sources with spectroscopic redshifts varies just between 21\% and 30\%. Obviously, spectra have not been sampled in a homogeneous manner over the three deep fields. In contrast to the spectroscopic redshifts, the completeness level of photometric redshifts is not only significantly higher, but also shows less fluctuation between the three fields (the fluctuations are at most 4\% at any given flux density threshold and at most 7\% between all different flux density cuts in the same deep field).

\begin{figure}
     \centering
     \includegraphics[width= 0.5 \textwidth]{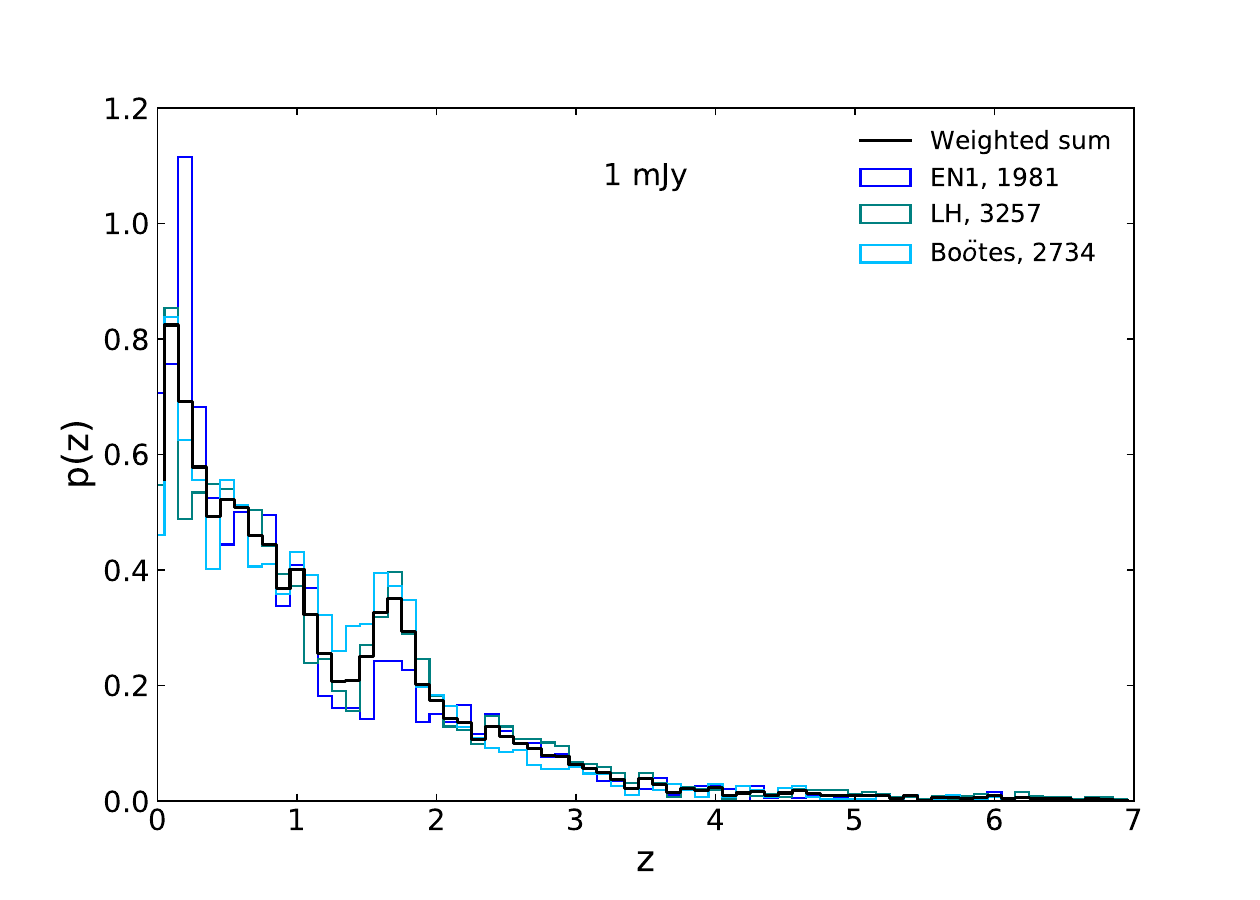}
     \caption{Weighted sum, $p(z)$, using \texttt{z\_best} value from the LoTSS Deep Fields for a flux threshold of 1 mJy. The numbers in the figures legend indicate the number of sources in each field used in the measurement. }
     \label{fig: pdf_zbest}
    \end{figure}

The photometric redshift estimates for radio sources come with varying uncertainties in measurements \citep{Benitez20, Brodwin_2006, Ken21}. 
As outlined above, the estimates are from a probability distribution function.
For the purposes of reducing the full redshift posterior into a single photometric redshift for catalogs, \citet{Ken21} define the single photometric redshift value, \texttt{z1\_median}, as the median of the primary peak in the $p_{i}(z)$ above the 80\% highest probability density credible interval \citep[HPD CI;][see also \citeauthor{Duncan2019}~\citeyear{Duncan2019} for a more detailed discussion]{Wittman2016}.
In the LoTSS-DF-DR1 release, the `best' redshift, \texttt{z\_best}, is then defined as the spectroscopic redshift if available, or \texttt{z1\_median} otherwise.
One could then compute the probabilistic distribution of all the sources as a function of redshift, $p(z)$, using these `best' estimate values for redshift, see Fig.~\ref{fig: pdf_zbest}. 
From this figure, the variation from field to field is evident, as well as likely unphysical features or biases arising from the sensitivity limitations within the multi-wavelength dataset (e.g. the peak in redshift distribution at $z\sim1.8$).

However, Fig.~\ref{fig: pdf_zbest} does not take into account the inherent uncertainties in the redshift measurement. In this work we therefore consider the full posterior pdf from photo-z estimates. 
Since the photometric redshift estimates are posterior pdfs that do incorporate all the available information about the redshift uncertainties, here we use a stacking approach to combine them. The advantage of such an approach is its simplicity, but it certainly makes strong implicit assumptions. The most important one that there is no correlation between the individual posterior pdfs, which is certainly not true as they all depend on the same systematic issues of a given set of multi-wavelength observations (see \citet{MalzHogg22} for a detailed discussion). Using estimates from three different fields with different multi-wavelength observations alleviates this problem, but does not solve it entirely.

In the following we consider the redshift distribution as a function of the flux density threshold. Here we provide more details on the procedure 
already presented and applied in \citet{Hale_2023} and \citet{Nakoneczny_2023}. Our procedure ensures that the stacked and weighted pdfs are properly normalised.
The posterior pdfs of each source are stored at $N_z = 701$ redshifts $z_i$ that are equally spaced between $z_0 =0$ and $z_{701}=7$. 
They are normalised to unity using a trapezoidal integration rule.
Our estimate of the redshift distribution in each field $f$ is then
\begin{equation}
\label{eq:1}
p_f(z) = \frac{1}{N_{f}} \sum_{s=1}^{N_f} p_{s}(z).
\end{equation}
Here, $p_{s}(z)$ is the posterior pdf for source $s$ in field $f$, $N_f$ denotes the total number of sources with posterior pdf in field $f$. 
We obtain posterior pdfs for flux density limited samples for 
each field and a weighted sum is taken by combining each of these $p_f(z)$, i.e.
\begin{equation}
\label{eq:2}
p(z) =  \sum_{f=1}^3 w_f p_f(z), \quad \sum_{f=1}^3 w_f = 1.
\end{equation}
Here, the weight $w_f$ represents the fraction 
of sources in a field~$f$. 

    \begin{figure*}
     \centering
     \includegraphics[width = 0.49 \textwidth]{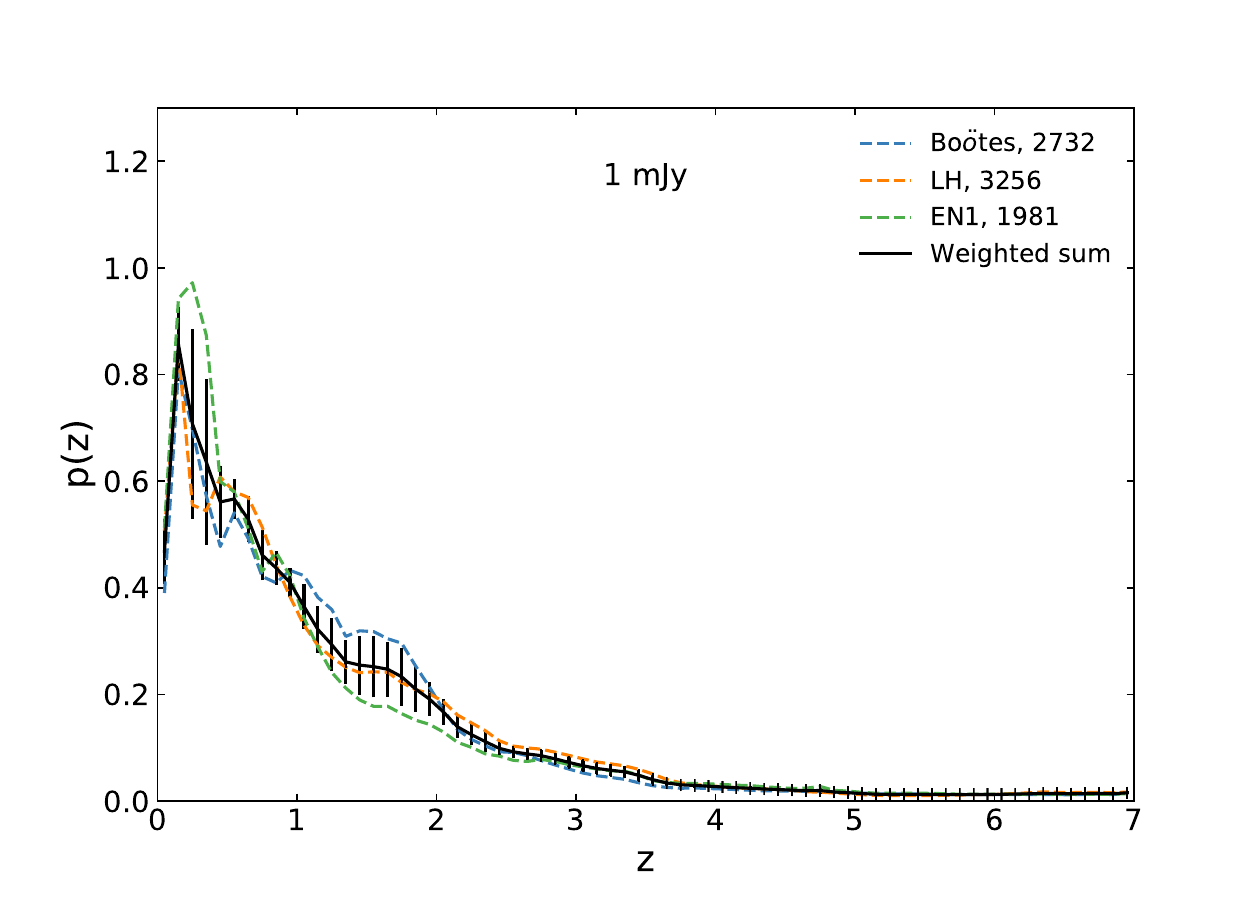}
     \includegraphics[width = 0.49 \textwidth]{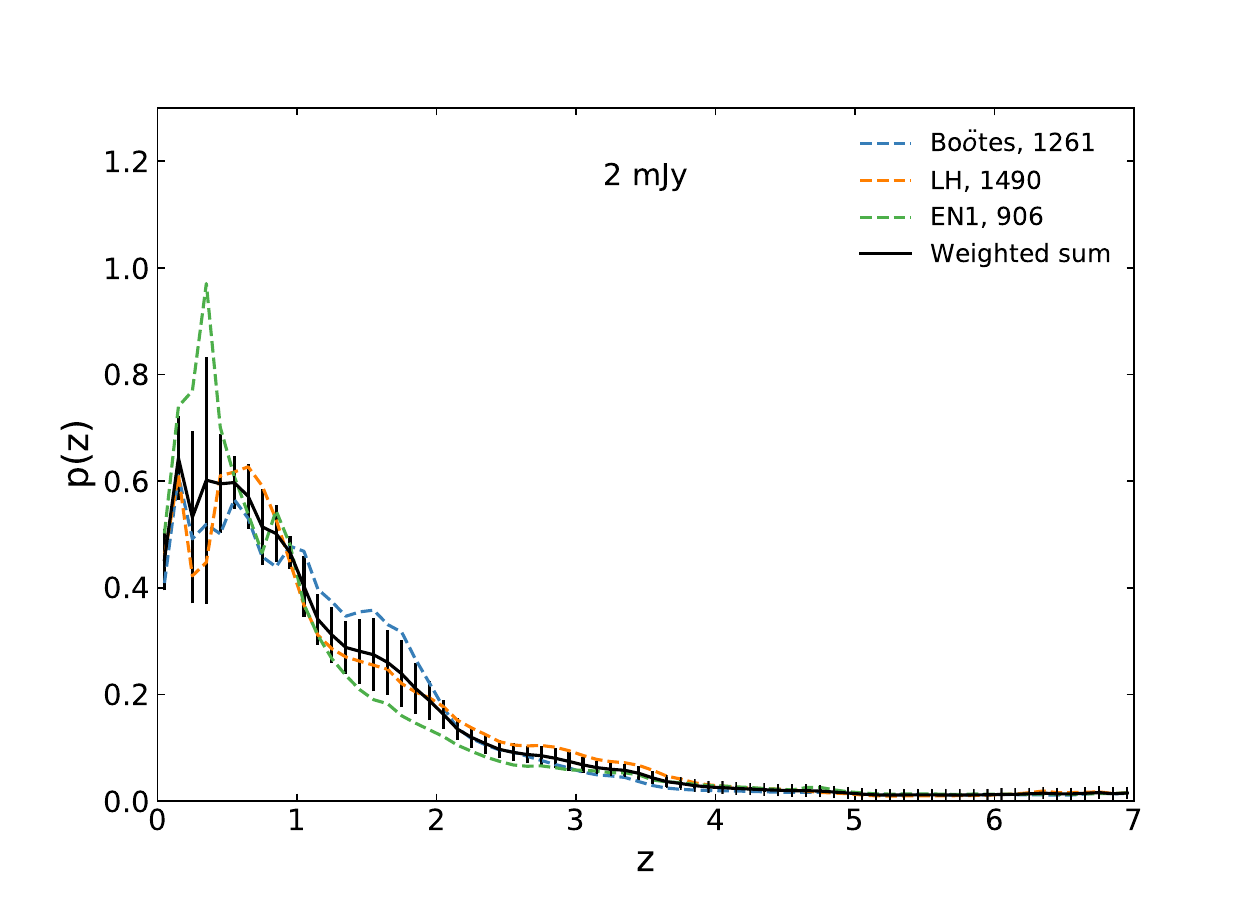}
     \includegraphics[width = 0.49 \textwidth]{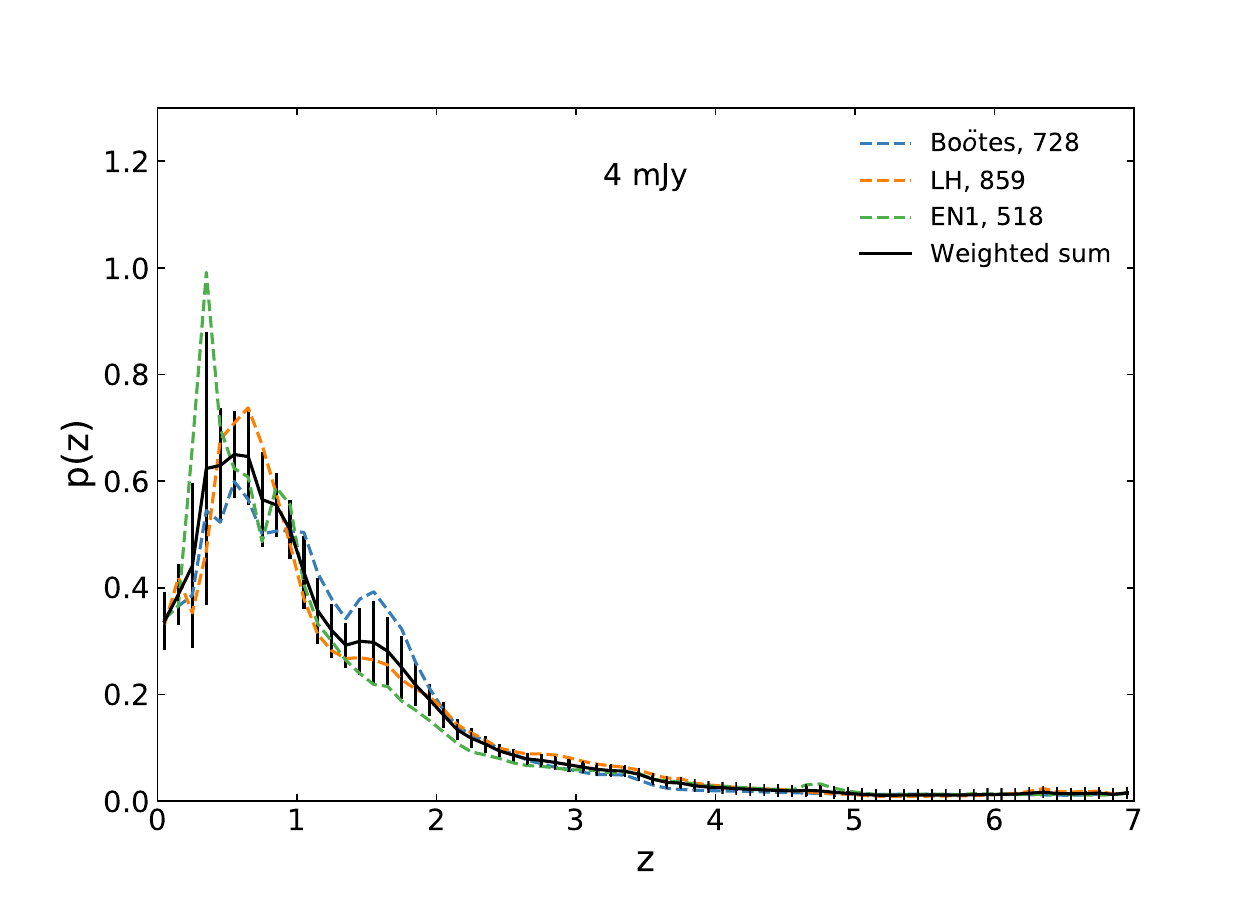} 
     \includegraphics[width = 0.49 \textwidth]{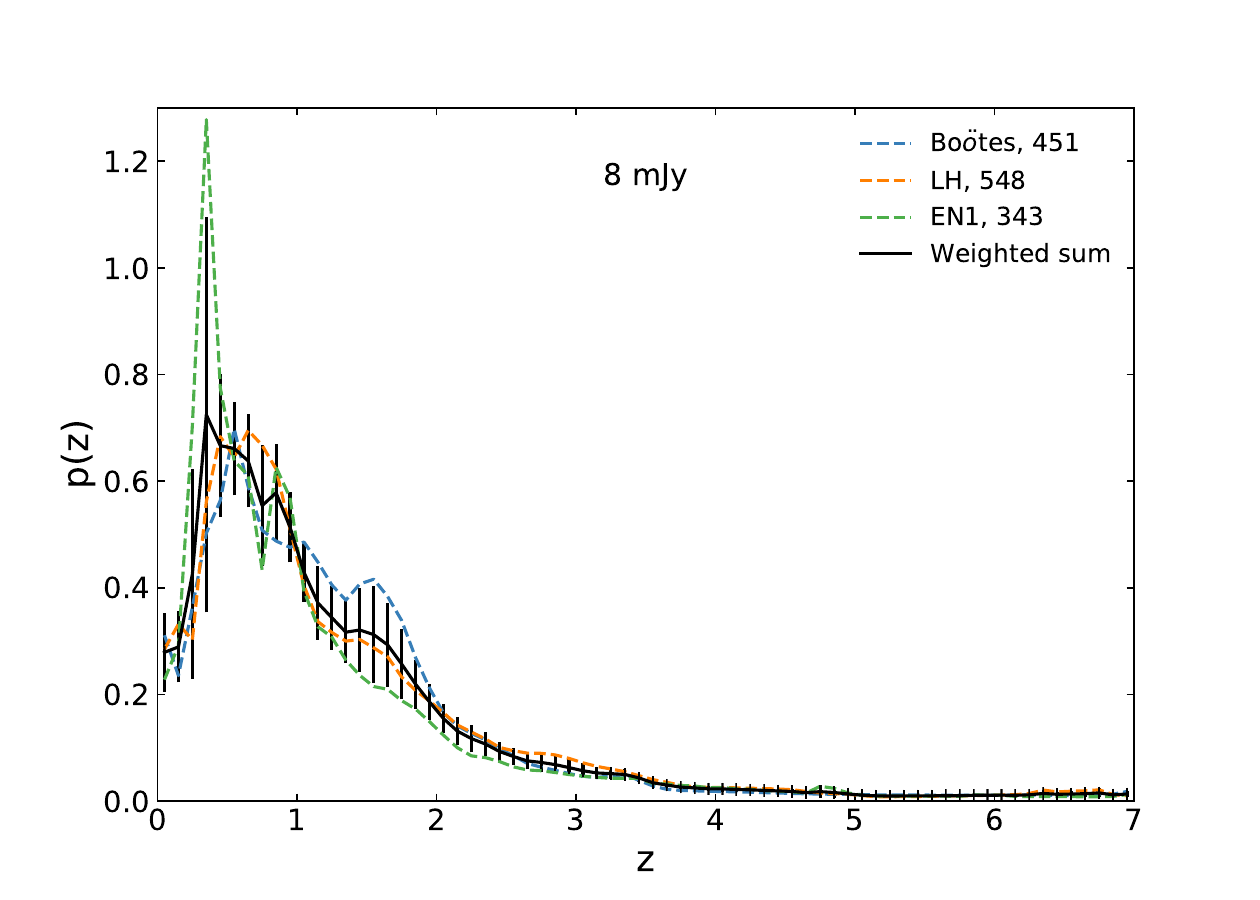}
     \caption{Redshift distribution for individual LoTSS Deep Fields, $p_f(z)$, and their weighted sum, $p(z)$, using photo-z pdfs from LoTSS Deep Fields DR1 for flux
     density thresholds of 1, 2, 4 and 8 mJy, respectively (top left to bottom right panels). Uncertainties are estimated by bootstrap resampling.}
     \label{fig: pz_1mJy}
    \end{figure*}

The errors on the weighted sum $p(z)$ are computed using the standard bootstrap resampling method. We make $N_\mathrm{b} = 50$ random samples for different flux density thresholds for each field; this is done by applying the method of selection with repetition, i.e.,\ a sample of $N_f$ sources is formed from the original catalogue by selecting sources randomly from the original catalogues for each field with repetitions allowed. The weighted sum is computed from each of these, resulting in a set of bootstrap samples, $\{ p_b(z) \}$, which are then used to compute uncertainties based on the empirical variance,
\begin{equation}
    \Delta p(z) = \sqrt {\frac {\sum_{b = 1}^{N_\mathrm{b}} [p_b(z) - \bar p(z)]^{2}}{N_\mathrm{b}-1}}. \label{eq:bootstrap}
\end{equation}
Here, $\bar p(z)$ is the weighted sum obtained from the means of the bootstrap samples of the three fields.

   \begin{figure}
     \centering
     \includegraphics[width= 0.5 \textwidth]{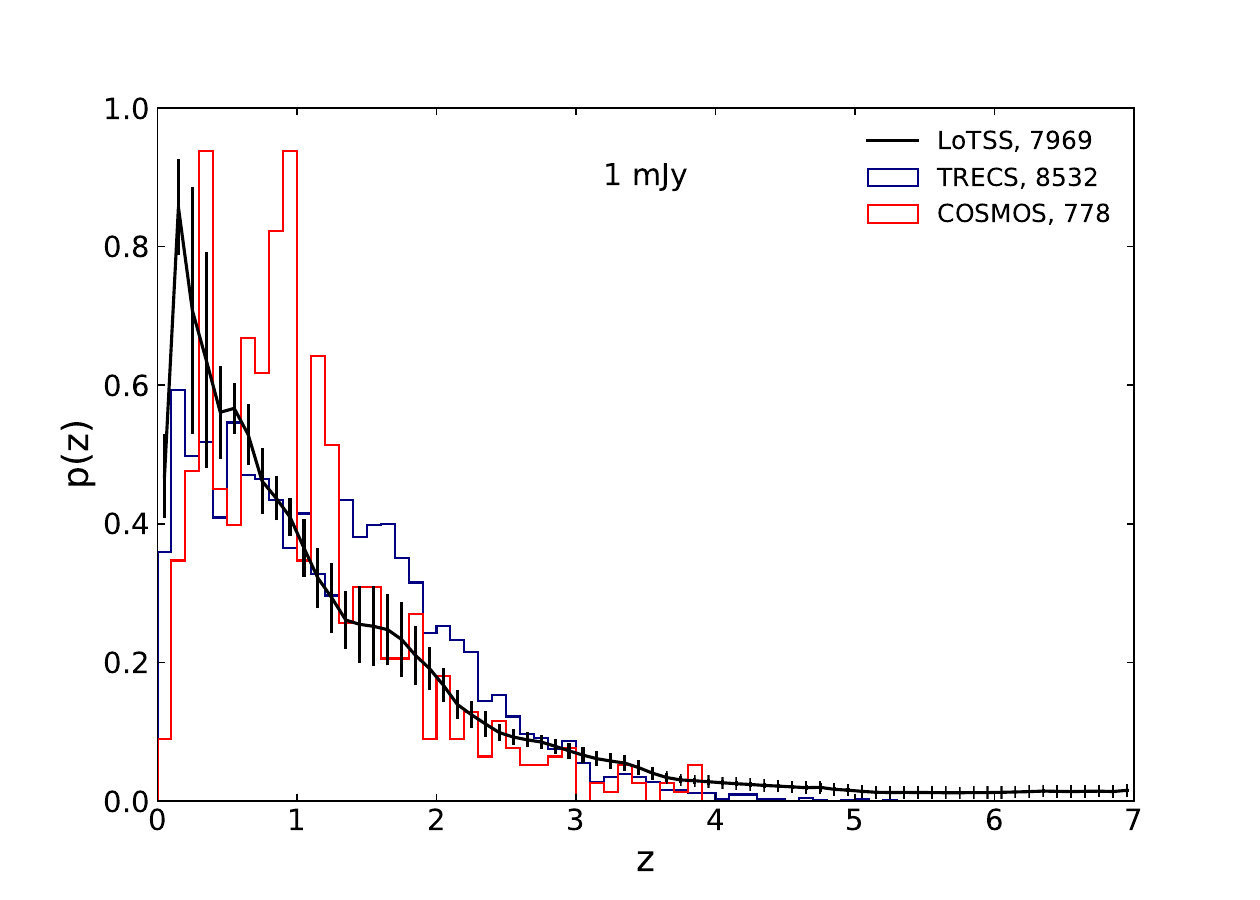}
     \caption{A comparison of $p(z)$ from T-RECS medium deep field and COSMOS $3$~GHz survey (scaled to 144 MHz) with LoTSS Deep Fields DR1 weighted sum $p(z)$ for a flux density threshold of $1$~mJy.}
     \label{fig: pz_trecs_lotss_cosmos}
    \end{figure}

Figure \ref{fig: pz_1mJy} presents the resulting $p_f(z)$ and $p(z)$ 
with an estimate of uncertainties for flux density thresholds of 
$1, 2, 4$ and $8$ mJy, respectively. We find reasonable agreement between all
three fields and at all flux density thresholds. Some notable variations are seen at $z < 0.5$, which are likely due to real differences in the large scale structure at low redshifts (the effect of cosmic variance), but might also be influenced by differences in multi-wavelengths coverage or effects of aperture photometry for nearby sources. Another notable variation is observed in the redshift range between 1 and 2. These differences at truly cosmological distance are most likely due to the different multi-wavelength coverage of the three deep fields. By employing errors based on the bootstrap resampling method, we 
capture both the systematic differences of photo-z measurements and cosmic variance. A comparison with Fig.~\ref{fig: pdf_zbest} shows that the pronounced peak in the redshift range 1 to 2 turns into an increased uncertainty on the pdfs in that range when the pdfs of individual sources are stacked.

In Fig.~\ref{fig: pz_trecs_lotss_cosmos}, we also compare our results to estimates of the redshift distribution for radio galaxies from COSMOS field \citep{refId0} --- after scaling the flux density $S \propto \nu^{\alpha}$ with an assumed spectral index of $\alpha = - 0.8$ and applying equivalent flux density thresholds, and the T-RECS simulation \citep{10.1093/mnras/stad1913}. We find only reasonable agreement of our weighted sum $p(z)$ with the $p(z)$ estimated from these two references. The T-RECS simulation shows an excess of sources at $1 < z < 2$ and a deficit at small redshifts compared to our results. The $p(z)$ estimated from the COSMOS field is in agreement with our weighted sum $p(z)$ except at around redshift values of 1. Note that the LoTSS Deep Field sample contains about an order of magnitude more sources than the COSMOS field at corresponding flux density thresholds and should therefore be less affected by cosmic variance.

\section{Angular two-point correlation from LoTSS-DR1}

    \begin{table*}
	\centering
    \caption{Best-fit values of the parameterisation $w(\theta) = A (\theta/1\ \mathrm{deg})^{1-\gamma}$ and integral constraint ($w_\Omega$) for the 
    LoTSS-DR1 value-added source catalogue after appropriate masking (mask 1) in the fit range $0.2$ deg $\leq \theta \leq 2$ deg and for four flux density thresholds. We also show the corresponding $r_0$ values for $\epsilon = \gamma -3$, $\epsilon = 0$ and $\epsilon = \gamma -1$. The best-fit values for $A$ and $\gamma$ for $1, 2$ and $4$ mJy were already reported in \citet{siewert}. All reported uncertainties are $68\%$ confidence intervals.
    We also report the median of the weighted posterior redshift distribution at the corresponding flux threshold, respectively.}
    \begin{tabular}{cccccccccc}
    	\hline\hline
    	\addlinespace[0.5ex] 
        $S_\mathrm{min}$  & $A$ & $\gamma - 1$& $w_\Omega$ &$\chi^2/$dof & $N$ & $r_0(\epsilon = \gamma -3)$ & $r_0(\epsilon = 0)$ & $r_0(\epsilon = \gamma -1)$ & $z_\mathrm{median}$ \\
        
        [mJy]   & $(\times 10^{-3})$ & & $(\times 10^{-3})$ & & & [$h^{-1}\mathrm{Mpc}$] & [$h^{-1}\mathrm{Mpc}$] & [$h^{-1}\mathrm{Mpc}$] \\\hline
		\addlinespace[1ex]  1   & $7.20^{+0.42}_{-0.42}$ & $0.68^{+0.08}_{-0.08}$&$1.9$  &5.78& 40\,599 
    & $10.9 \pm 2.8$ & $14.2 \pm 2.5$ & $15.8 \pm 2.9$ & 0.92 \\
		\addlinespace[0.5ex] 2  & $5.11^{+0.59}_{-0.60}$ & $0.74^{+0.16}_{-0.16}$&$1.2$ &2.70& 19\,719 
    & $10.1 \pm 2.6$ & $13.3 \pm 2.1$ & $15.1 \pm 2.5$ & 0.96 \\
		\addlinespace[0.5ex] 4  & $7.45^{+0.95}_{-0.95}$ & $0.46^{+0.21}_{-0.20}$&$3.0$ &2.34& 11\,269 
    & $9.5 \pm 2.8$ & $16.0 \pm 3.1$ & $18.3 \pm 3.9$ & 0.99 \\
		\addlinespace[0.5ex] 8  &$7.69^{+0.33}_{-0.33}$ &$0.89^{+0.49}_{-0.49}$ & 1.4 &1.86 & \ 3\,430 
   & $13.3\pm 5.2$ & $17.6 \pm 6.1$ & $20.8 \pm 7.4$ & 0.99 \\
         \addlinespace[1ex]\hline 
    \end{tabular}
    \label{tab:BestFit}
    \end{table*}

    The angular two-point correlation function, $w(\theta)$, quantifies the angular clustering of extragalactic sources \citep{Peebles1980}. 
    It measures the excess probability of finding a source in the vicinity of another source, separated by an angle $\theta$. In case of Poisson distributed point sources this function would be zero. 
    In this work we parameterise $w(\theta)$ by a simple power law,
    \begin{equation}
        w(\theta)= A \displaystyle \left(\frac{\theta}{1\, \mathrm{deg}}\right)^{1 - \gamma}.
    \end{equation}
    This ansatz is motivated by a corresponding power-law ansatz for the three dimensional spatial correlation function, see the next section for more details. 
    
\begin{figure}
     \centering
     \includegraphics[width= 0.5 \textwidth]{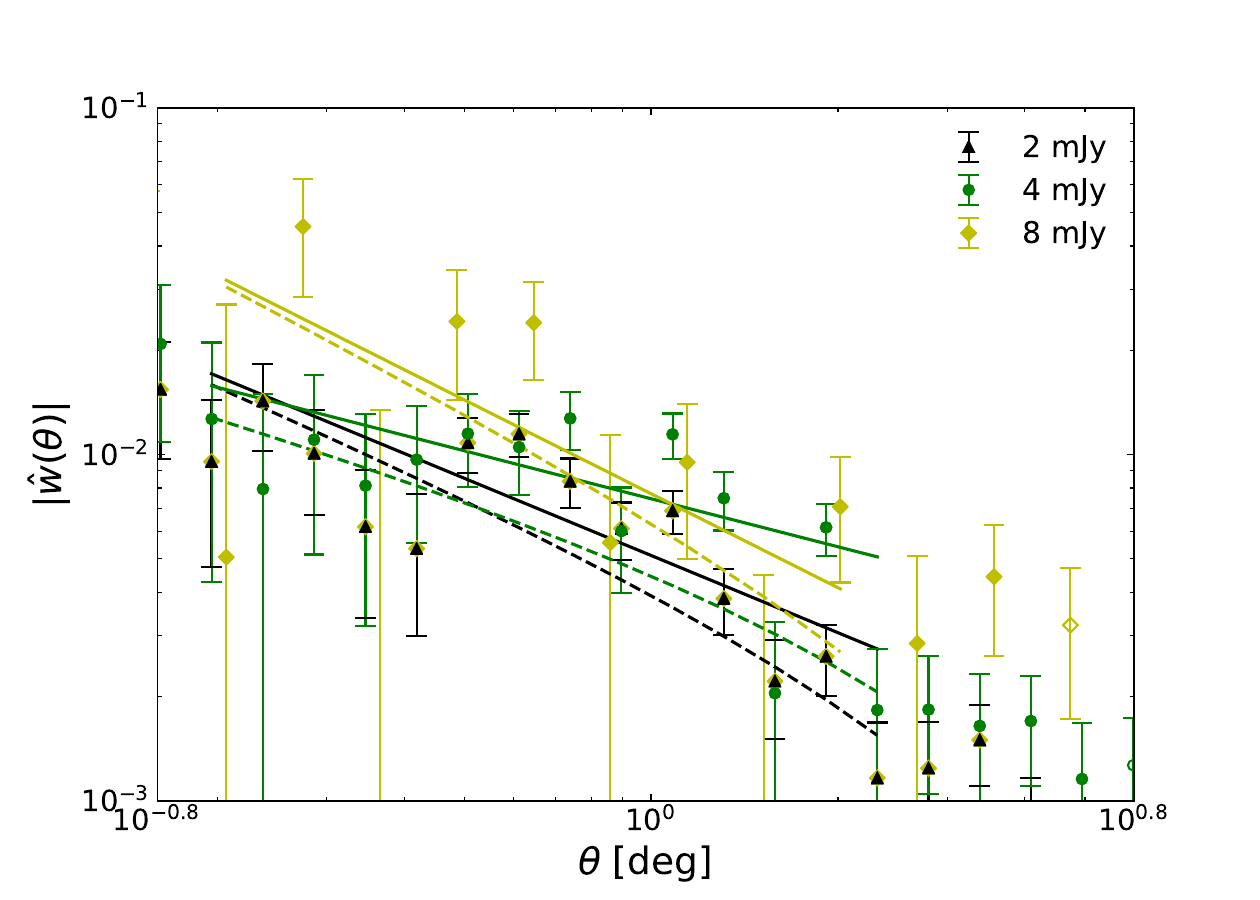}
     \caption{Absolute value of two-point angular correlation function for different fluxes and their corresponding fit lines (solid lines show the best-fit power-law and dashed lines show the fits after including the effect of the integral constraint) over the fit-range $0.2$ deg $ < \theta < 2$ deg with same colour codes. Open symbols for data points indicate negative values of $w(\theta)$. Data and fits for 8 mJy are new, for 2 and 4 mJy we take them from \cite{siewert}.}
     \label{fig: w_theta_fits}
    \end{figure}

    \citet{siewert} measured the angular two-point correlation function and fitted the amplitude of angular clustering $A$ and the index $\gamma$ for radio sources above different flux density thresholds LoTSS-DR1 radio sources from the HETDEX spring field \citep{Shimwell19}. The basis for the angular clustering measurements is the LoTSS-DR1 value added source catalogue \citep{Williams2019} containing 318\,520 radio sources observed over 424 square degrees. 
    The measurements of $w(\theta)$ make use of the optimal estimator originally defined by \citet{LandySzalay1993}. Measurements of $w(\theta)$ from \citet{siewert} are reproduced in Fig.~\ref{fig: w_theta_fits}, where we also add a new measurement at $S_\mathrm{min} = 8$ mJy. The estimated angular correlation, $\hat w (\theta)$, is biased due to the integral constraint which arises due to the finite geometry of the survey.
    Therefore, we also report the estimated bias $w_{\Omega}$, that we obtain from an iterative fit, i.e.\ $\hat w(\theta) = A (\theta/1 \mathrm{deg})^{1-\gamma} - w_\Omega$ (see appendix of \citeauthor{siewert} \citeyear{siewert} for details). The fit range was chosen in \citet{siewert} as $0.2$ deg $< \theta < 2$ deg, avoiding the effects of non-linear structures at the small scales and systematic flux density uncertainties between different pointings on the larger scales (the typical distance between two pointings is $1.7$ deg). We consider the `mask 1' measurement for flux density thresholds of $1, 2$ and $4$ mJy and the additional flux density threshold of 8 mJy, for which we follow the same analysis pipeline, but use only half the number of bins for the angular separation $\theta$ to retain a similar number of source pairs per bin. Fig.\ \ref{fig: w_theta_fits} presents the results for the $2, 4, 8$ mJy flux density thresholds. 
    We summarise the measurements from \citet{siewert} and our new results in Table~\ref{tab:BestFit}. We also quote the goodness of fit and number of radio sources after masking the survey area and flux density cut. We obtain the best goodness-of-fit for the $8$~mJy sample, $A = 7.69 \pm 0.33$ and $\gamma - 1 = 0.89 \pm 0.49$. However, the rather small sample size leads to rather large uncertainies. The smallest uncertainties with a still acceptable goodness-of-fit are found for the $2$~mJy sample, $A= 5.11 \pm 0.60$ and $\gamma - 1 = 0.74 \pm 0.16$, as discussed in \citet{siewert}.

    These measurements of $A$ (reported at $1$ deg) 
    are higher to measurements in the literature:
    For NVSS, $A$= $(1.45 \pm 0.15) \times 10^{-3}$, $\gamma -1 = 1.05 \pm 0.10$ has been obtained by \citet{Blake2004} for $S_{1.4\, \mathrm{GHz}}> 10$ mJy and $A$= $(1.0 \pm 0.2) \times 10^{-3}$ for $\gamma -1 = 0.8$ by \citet{Overzier2003} for similar flux density thresholds. 
    In a more recent study \citet{RanaBagla2019, RanaBaglaErratum} have measured 
    $A$= $(8.4 \pm 0.5) \times 10^{-3}$ and $\gamma -1 = 0.77 \pm 0.15$ for TGSS at $S_{154\, \mathrm{MHz}} >100$ mJy, which is in better agreement with our measurements, especially at our highest flux density threshold of $8$ mJy. Note that \citet{siewert} concluded that only LoTSS-DR1 results at and above $2$ mJy should be used for cosmological analysis, as sources at lower flux densities still suffer from systematic issues that have not been understood well enough in LoTSS-DR1. Discussions of these potential systematic issues are commented on in \citet{Hale_2023}. The newly added data point for flux densities above $8$ mJy is in agreement with the results at $2$ and $4$ mJy, however the significantly reduced number of radio sources at flux densities above $8$ mJy reduces the statistical significance of the measurement. In \citet{Hale_2023} the LoTSS-DR2 angular two-point correlation has been measured for a flux density thereshold of $1.5$ mJy, where for a fixed value of $\gamma - 1 = 0.8$, an amplitude of $A = (2.88^{+0.07}_{-0.06}) \times 10^{-3}$ was found.\footnote{For the reader's convenience we convert 
    $\log_{10} A = -2.54^{+0.01}_{-0.01}$ obtained for the fit-range $0.03 \deg \theta < 1 \deg$ by \citet{Hale_2023}.}
    
\section{Clustering scale}

    Ignoring relativistic effects \citep{Yoo10, ChallinorLewis11, BonvinDurrer11}, the relation between the typical comoving length scale of clustering, $r_0$, and the angular two-point correlation $w(\theta)$ 
    is given by Limber's equation \citep{Limber53}.
    One has to assume the statistical isotropy and homogeneity of the large-scale structure and a functional form for the spatial two-point correlation function, $\xi(r)$, where $r$ denotes the comoving distance. Often, a power law is assumed, $\xi(r)= (r/r_0)^{-\gamma}$, with $\gamma > 0$, see e.g.\ \citet{Peebles1980}. Increasing the clustering length $r_0$, implies an increasing correlation of any pair of objects at fixed distance $r$. This ansatz ignores the evolution of large scale structure. To take the evolution of galaxy clustering into account, a dependence on redshift must be introduced. A simple ansatz is to model this evolution as a power of $1+z$,
    \begin{equation}
        \xi(r,z)= \left(\frac{r}{r_0} \right)^{-\gamma} (1+z)^{\gamma -3 -\epsilon}, 
    \end{equation}
    where $\epsilon$ parameterises the type of the galaxy clustering model \citep{1977ApJ...217..385G, Overzier2003}. $\epsilon \ = \ 0$ describes the \textit{stable clustering model}, which assumes that cosmic structures are gravitationally bound at small scales and do not evolve over the observed range in redshift. In this model, galaxy clusters neither expand nor contract with the Universe and have a correlation function which decreases with
    redshift; $\epsilon \ = \gamma - 3$ parameterises the \textit{comoving clustering model}, in which the
    large scale structures expand with the Universe and hence their correlation function remains fixed in comoving coordinates. In that model, cosmic large-scale structures would not (yet) be gravitationally bound. Finally, $\epsilon \ = \gamma - 1$ parameterises the \textit{linear growth model}, in which the clustering is described as per the linear perturbation theory (before the cosmological constant starts to dominate). However, 
    the evolution of clustering properties is degenerate with the evolution of the galaxy clustering bias (the effect that the radio source density does not necessarily trace mass density).
    
    We first state the result of Limber's approximation, which holds for small angular scales and a Universe dominated by non-relativistic matter (see \citeauthor{Simon2007} \citeyear{Simon2007} for a detailed discussion of the range of validity of the approximation)
    \begin{equation}
    w(\theta)= r_{0}^{\gamma} \sqrt{\pi}\frac{\Gamma[(\gamma -1)/2]}{\Gamma[\gamma/2]} \theta^{1-\gamma}\! \!\! \int_{0}^{\infty}\!\!\! d\bar{r} \tilde{p}^2(\bar{r}) 
    \bar{r}^{1-\gamma}
    \left[ 1 + z(\bar{r})\right]^{\gamma - 3 - \epsilon}\!\! ,
    \label{eq: limber} 
    \end{equation}
    which we can write as
    $w(\theta)=A(r_0) \,  (\theta/1 \, \mathrm{deg})^{1-\gamma}$. Above, $\Gamma[x]$ denotes the Gamma function.
    Based on a measurement of $A$ and $\gamma$, this relation can be used to compute $r_0$ when a model for the evolution of galaxy clustering is assumed. In Eq.~(\ref{eq: limber}) $\bar{r}$ is the mean comoving radial distance of two sources separated by comoving distance $r$. The mean comoving distance corresponds to a redshift of $z = z(\bar r)$ and $\tilde p(\bar r)$ is the window function or pdf in comoving distance space for, in our case, the radio sources in the LoTSS deep fields.
    From observations one generally measures the window functions as a function of redshift $p(z)$. Equation~(\ref{eq: limber}) needs to be modified accordingly. For this we need a relation between the comoving radial distance $\bar r$ and redshift $z$, for which one has to assume a cosmological model.
    
    We consider the case where distances are given by a spatially isotropic, homogeneous metric and assume a flat LCDM model. The radial line-of-sight comoving distance for a flat Universe is then given by,
    \begin{equation}
    r(z) = \displaystyle \frac{c}{ H_0} \int_{0}^{z} \displaystyle \frac{dz'}{E(z')},
    \label{eq: rz}
    \end{equation}
    where 
    \begin{equation}
    E(z) = \sqrt{ \Omega_\mathrm{M}(1+z)^{3} + 1 - \Omega_\mathrm{M}},
    \label{eq: Ez}
    \end{equation}
    with $\Omega_{\mathrm{M}}$ 
    denoting the dimensionless matter density of the present Universe and today's Hubble rate $H_0 = 100\ h$ km/s/Mpc.
      
    From the normalisation condition for the redshift distribution, $\int_0^\infty p(z)\, \mathrm{d} z = 1$, we find 
    \begin{equation}
        \tilde p(r(z)) = \frac{H_0 E(z)}{c} p(z).
    \end{equation}
    One can then re-write the integral term in relation (\ref{eq: limber}) in terms of redshift $z$ as,
    \begin{equation}
    \begin{split}
    I(\gamma,\epsilon, \Omega_\mathrm{M}) &= \left(\frac{H_0}{c}\right)^{-\gamma} \int_{0}^{\infty} \mathrm{d}\bar{r} \tilde{p}^2(\bar{r}) \bar{r}^{1 - \gamma} (1+z)^{\gamma - 3- \epsilon} \\
    &= \displaystyle  \int_{0}^{\infty} \mathrm{d} z E(z) p^2(z) (1+z)^{\gamma - 3- \epsilon} \left(\int_{0}^{z} \frac{dz'}{E(z')}\right)^{1-\gamma}.
    \end{split}
    \end{equation}
    From this expression, given the values for the density parameters, an evolution model of galaxy clustering, and an observed redshift distribution of galaxies, one can easily compute the clustering length $r_0$. Its unit follows from the unit of the Hubble distance $c/H_0$, which is $3000\ h^{-1} $Mpc. 
    We obtain
    \begin{equation}
    \begin{aligned}
    A = &  \sqrt{\pi}  \frac{\Gamma[(\gamma-1)/2]}
    {\Gamma[\gamma/2]} I(\gamma, \epsilon, \Omega_\mathrm{M}) \left( \frac{ r_0H_0}{c}\right)^{\gamma} 
    \left(\frac{\pi}{180}\right)^{1-\gamma},
    \end{aligned}
    \end{equation}
    when we measure the angular separation in units of degrees. 
    Thus the measured strength of the angular correlation depends on the cosmological model ($H_0, \Omega_\mathrm{M}$), the correlations length $r_0$, the exponent $\gamma$, as well as the evolution of clustering, described by $\epsilon$ of the given probe. Obviously, the cosmological model as well as the clustering properties of dark and baryonic matter should not depend on the flux density of the chosen sample. However, the specifics of the chosen probe might depend on the flux density cut and differ in clustering evolution.

    \begin{figure}
     \centering 
     \includegraphics[width= 0.5 \textwidth]{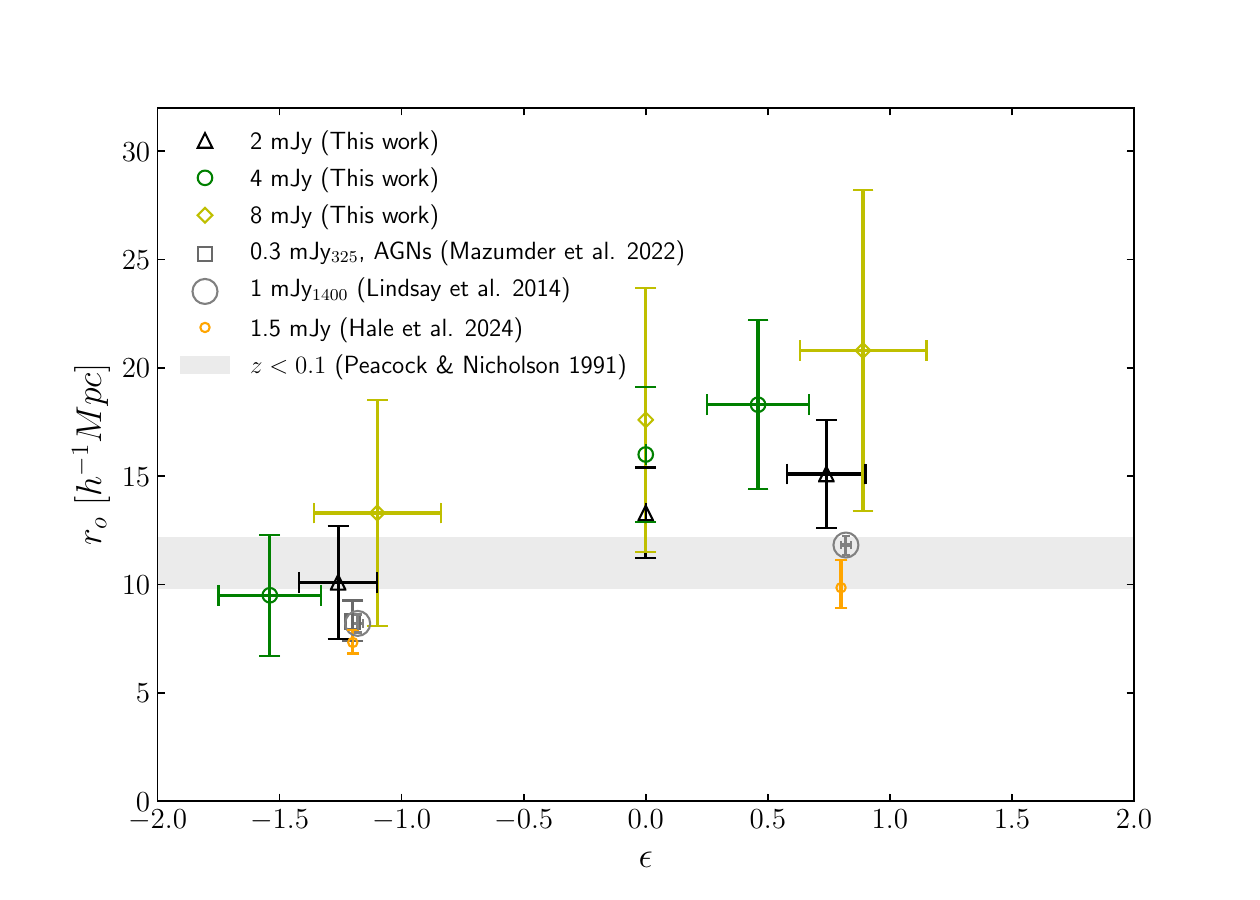}
     \caption{Comoving correlation length for different flux density thresholds at the central LoTSS frequency of 144 MHz and clustering parameter $\epsilon$. For comparison a selection of measurements from the literature are shown. The result of \citet{1991MNRAS.253..307P} uses bright radio galaxies at $z < 0.1$ and does therefore not depend on the clustering evolution (ignoring any factors of $1+z$, they implicitly use the comoving clustering model), thus applies to any value of $\epsilon$. The analysis of \citet{Mazumder2022} uses observations at $325$ MHz and distinguishes AGNs and SFGs. We show their results for AGNs. The radio sources from \citet{Lindsay2014} (FIRST at $1.4$ GHz) and \citet{ Hale_2023} (LoTSS-DR2 at 144 MHz) are mostly AGNs. 
     }
     \label{fig: ro_eps}
    \end{figure}
      
For different flux thresholds and value of $\epsilon$ we compute the clustering length. 
The expectation and uncertainty of $r_0$ are calculated from the corresponding redshift distribution and the values of $A$ and $\gamma$ from the wide field LoTSS-DR1 as mentioned in Table~2.  
We construct a multivariate normal distribution using these values and the associated covariance matrix, to randomly select ($A$,$\gamma$) pairs from such a distribution. 
We then randomly select 100 distributions $p_\mathrm{b}(z)$, as inferred from LoTSS-DF-DR1 
(see Sect. 2), and draw 1\,000 ($A$,$\gamma$) pairs for each of those $p_\mathrm{b}(z)$. 
Thus we sample 100\,000 $r_0$ values for each case.
Then, similar to the bootstrap error computation as described in Eq.~(\ref{eq:bootstrap}) we compute expectation (mean) and uncertainty of $r_0$. 
The results are shown in Fig.~\ref{fig: ro_eps} and in Table~\ref{tab:BestFit}. 
We also show the median redshift of the four flux density limited redshift distributions from LoTSS-DF-DR1, which are between 0.9 and 1. 
    
As shown in Table~\ref{tab:BestFit}, the most precise measurement of $r_0$ is obtained 
for 2 mJy with $(10.1 \pm 2.6) h^{-1}$Mpc, $(13.3 \pm 2.1)h^{-1}$Mpc, and $(15.1 \pm 2.5) h^{-1}$Mpc for 
$\epsilon = \gamma - 3, 0$, and $\gamma -1$, respectively. The clustering length increases with the value of $\epsilon$. We also find good agreement between the different flux thresholds for all three models of clustering evolution. For the comoving clustering model ($\epsilon = \gamma - 3$) we find $(9.5 \pm 2.8) h^{-1}$Mpc and $(13.3 \pm 5.2) h^{-1}$Mpc, for 4 mJy and 8 mJy, respectively.
In principle an inconsistent clustering model could be identified by wildly inconsistent clustering lengths for different flux density cuts, but that is apparently not the case here. However, we should not expect them to be identical, as the different flux density samples contain a different mix of AGNs and star forming galaxies (SFGs) (see Best et al. 2023), with almost all radio sources above 8 mJy being AGNs and an increasing (but small) fraction of SFGs as we lower the flux threshold. This is in line with the slight decrease of medium redshift with decreasing flux density threshold, as the first SFGs that are detected tend to be at smaller redshifts. 
    
Several studies have measured the comoving clustering length for galaxies, including radio galaxies \citep[see, e.g.][]{2022A&ARv..30....6M}. The vast literature on that subject suggests that $r_0$ varies depending on galaxy properties and environment. Especially luminous and old galaxies give rise to larger clustering length, 
in contrast to less luminous or blue galaxies.
For nearby radio galaxies ($z < 0.1$) \citet{1991MNRAS.253..307P} reported a clustering length of $r_0 = (11.0 \pm 1.2)\ h^{-1}$Mpc (as this measurement summarises the amount of clustering today, it can serve as a reference point for all three evolution models). Later \citet{Overzier2003} reported a value of $r_0 = (14 \pm 3)\ h^{-1}$Mpc for an analysis of Faranoff-Riley type II (FRII) radio galaxies, which contribute dominantly to the correlation length at high flux densities ($S_{1.4 \mathrm{GHz}} > 200$ mJy) in NVSS and FIRST. 
At lower flux densities and for the mix of all radio sources, they find smaller clustering length of $r_0 = (4$ to $10)\ h^{-1}$Mpc, depending on the flux density threshold. Based on a simulated redshift distribution of FIRST sources at 
$S_{1.4\, \mathrm{GHz}} > 1$ mJy, \citet{Lindsay2014} find $r_0 = 8.20^{+0.41}_{-0.42}\ h^{-1}$Mpc (scaling with a spectral index of $\alpha = - 0.8$ or $-0.7$ to the central LoTSS frequency we should compare to $S > 6.2$~mJy or $4.9$~mJy, respectively). 
These numbers correspond to the analysis for the \textit{comoving clustering model} ($\epsilon = \gamma -3$).
Those findings have been confirmed by more recent studies \citep[see, e.g.][]{2022A&ARv..30....6M}.

Inspecting our results for the \textit{comoving clustering model} (see Fig.~\ref{fig: ro_eps} and Table~\ref{tab:BestFit}), we find that they are in good agreement among each other for all considered flux density thresholds and with the values that have been measured for radio galaxies at $1.4$ GHz (see above). Our results are consistent with the trend observed at $1.4$ GHz that increasing flux density thresholds seem to go along with stronger clustering. 
Combining our findings with results based on TGSS \citep{Dolfi2019,RanaBagla2019,RanaBaglaErratum}, 
we see the same trend. 
In \citet{Hale_2023} the analysis of LoTSS-DR2 reveals for $\gamma - 1 = 0.8$ a value of 
$r_0 = (7.32^{+0.59}_{-0.51})\ h^{-1}$Mpc
at $1.5$ mJy 
(note the different fit range: $0.03 \deg < \theta < 1 \deg$),
also consistent with this trend. 
At 325 MHz, \citet{Mazumder2022} measured the clustering length from a deep observation of the LH for AGNs (with $z_\mathrm{median} = 1.02$, very similar to our analysis) and SFGs separately at a flux density above 0.3 mJy, corresponding to about 0.6 mJy at the LoTSS frequencies. They find $r_0^\mathrm{AGN} = 8.30^{+0.96}_{-0.91}\ h^{-1}$Mpc, when assuming $\gamma - 1 = 0.8$. Our result is close to their finding. 

Let us also investigate the \textit{stable clustering} ($\epsilon = 0$) and \textit{linear growth model} ($\epsilon = \gamma - 1$), with results for both of them presented in Fig.~\ref{fig: ro_eps} and Table~\ref{tab:BestFit}. We measure values for $r_0$ that are close to those measured for galaxy clusters (e.g.\  $(24 \pm 9)\ h^{-1}$~Mpc from \citealt{Bahcall,Postman}). They show the same trends of more clustering for brighter objects, but provide clustering lengths that exceed the local reference measurement from \citet{1991MNRAS.253..307P}, which makes them 
less plausible, as our values probe the strength of clustering at a median redshift close to 
unity, and thus both models would suggest that the clustering of radio galaxies actually decreased since redshift of unity. While in principle this is expected for a LCDM model in the future (all not gravitationally bound structures will be diluted in the de Sitter future of the universe), the onset of acceleration is not far enough in the past to make those models plausible. It is interesting to note that the for models with larger $\epsilon$ the trend of more clustering for higher flux density threshold is more pronounced than for the comoving model.

Looking at our findings and measurements from the literature, the \textit{comoving clustering model} can easily match the data for the local and radio loud AGNs, which implies that in fact there is no redshift dependence in $\xi(r)$. This would mean that 
the galaxy clustering bias must be a function of redshift inversely proportional to the growth of the large scale structure. Indeed \citet{Hale_2023} and \citet{Nakoneczny_2023} find that such a bias model provides a better description of the LoTSS-DR2 data, compared to a redshift independent galaxy clustering bias.

Finally, we investigate the effect that using the improved posterior redshift distribution in equation (\ref{eq:1}), and shown in Fig.~\ref{fig: pz_1mJy}, has over obtaining the redshift distribution based on catalogued $z_\textrm{best}$ values, for which an example is shown in Fig.~\ref{fig: pdf_zbest}. For the 
flux density threshold with the best statistics but still above the systematic limitations of LoTSS-DR1, namely $S > 2$ mJy, 
we find $r_0 =  (9.9 \pm 2.4)\ h^{-1}$ Mpc,  $(12.9 \pm 1.9)\ h^{-1}$ Mpc, and $(14.3 \pm 2.3)\ h^{-1}$ Mpc
for $\epsilon = \gamma - 3, 0$, and $\gamma -1$, respectively. Those are less than $1 \sigma$ smaller 
than the correlation lengths measured by means of the full posterior distribution $p(z)$ and shown in 
Fig.~\ref{fig: ro_eps} and Table~\ref{tab:BestFit}. For these estimates only the uncertainties in 
$(A,\gamma)$ are taken into account.
The correlation lengths based on $z_\textrm{best}$ values tend to underestimates the high-z tail of the distribution, as AGNs are sometimes mistaken for SFGs at lower redshift, but the opposite happens less likely, as was shown by \cite{Ken21}. Our method accounts for those systematic and thus results in a slightly larger value of the clustering length as moving objects at fixed angular distance to larger redshift increases their physical distance.
Our findings seem to indicate that the measurement of the correlation length is not strongly depending on the assumptions made here and for a rather limited survey area of LoTSS-DR1 with its large uncertainties on $A$ and $\gamma$. However, already with the significantly reduced statistical uncertainties of LoTSS-DR2, see \citet{Hale_2023} who report uncertainties of $\Delta r_0 \approx 0.6\, h^{-1}$ Mpc, this is no longer the case. 

\section{Conclusions}
    
In this work we obtained estimates of the redshift distribution $p(z)$ of LoTSS Deep Field radio sources 
for various flux density limits and quantified their uncertainties $\Delta p(z)$ by means of bootstrap 
resampling. We based our estimates on stacking the posterior pdfs for individual radio sources as described 
and determined in \citet{Ken21}, which had made use of the good multi-wavelength coverage 
\citep{Kondapally2021} of radio sources from three LoTSS Deep Fields \citep{Tasse2021,Sabater2021}. These had 
allowed \citet{Ken21} to obtain posterior pdfs for the photometric redshift of $96\%$ of all radio 
sources in the survey region with good multi-wavelength information. After applying a flux density threshold, 
the photometric redshift completeness drops to $91\%$. We have implicitly assumed that the remaining $4\%$ to 
$9\%$ of radio sources follow the same distribution, which might result in an underestimation of the number of 
radio sources at redshifts above unity. By averaging over three different deep fields and over a total area of 
about 26 square degrees, we reduce cosmic variance and estimate the effects of systematic issues
by bootstrap resampling of the data. We conclude that LoTSS radio sources above a flux density of $1$ mJy 
have a median redshift of about unity.

We also used the inferred redshift distribution of the LoTSS deep fields and applied it on 
the wide field clustering data from LoTSS-DR1 for three different flux density thresholds, 
$2, 4$ and $8$ mJy (the results for $1$ mJy are shown for completeness, but should not 
be trusted as they still suffer from systematic issues; see also the discussions in 
\citealt{siewert} and \citealt{Hale_2023}). We find good consistency for the 
\textit{comoving clustering model} in which the clustering structures have formed 
well before they are observed and probed by the survey, which indicates 
that halos hosting radio galaxies (which are rare compared to normal optical or 
infrared galaxies) are in most cases not gravitationally bound to each other and 
therefore expand with the Hubble flow. Our most precise result is that 
for $2$ mJy, with a clustering length of $r_0 = (10.1 \pm 2.6)\ h^{-1}$Mpc. Our bootstrap analysis 
shows that the precision in this analysis is largely limited  by the precision of the measured 
angular two-point correlation function rather than the redshift distribution of sources. This will 
change with larger LoTSS samples and significantly improved measurements for $A$ and $\gamma$, as already clear from the LoTSS-DR2 analysis in \citet{Hale_2023}.

Our study of the redshift distribution of radio sources also allows us to estimate the size of the sampled comoving volume of the wide area LoTSS. As an estimate we weight the comoving volume of the Universe by the redshift distribution of LoTSS radio sources,
\begin{equation}
V_\mathrm{LoTSS} = \int \mathrm{d} \Omega \int \mathrm{d} z \frac{p(z)r^2(z)}{H(z)}.
\end{equation}
Thus, LoTSS-DR2 \citep{Shimwell2022}, which covers about 1/8 of the full sky, probes $\approx 3.3 h^{-3}$\ Gpc$^3$. 
After LOFAR observing cycle 20 (which finishes in summer 2024) the
coverage of LoTSS will allow us to 
increase this volume to about $\approx 10 h^{-3}$~Gpc$^3$. This demonstrates the unique potential to measure the largest cosmic structure of combining wide area radio continuum surveys with multi-wavelength information. The WEAVE-LOFAR survey \citep{2016sf2a.conf..271S} aims at obtaining spectroscopic redshifts for all LoTSS sources above a flux density of 8 mJy. The clustering study of this work thus serves as a first reference point for much more detailed studies of the three dimensional large scale clustering based on radio selected spectroscopic redshifts, which, compared to optically and infrared selected samples, will provide an independent and complementary probe of the Universe at the largest scales.

\begin{acknowledgements}
We thank Maciej Bilicki for discussions and comments. NB and DJS acknowledge financial support by Deutsche Forschungsgemeinschaft (DFG) under grant RTG-1620 `Models of Gravity'. CLH acknowledges support from the Leverhulme Trust through an Early Career Research Fellowship and from the Hintze Family Charitable Foundation through the Oxford Hintze Centre for Astrophysical Surveys. CSH's work is funded by the Volkswagen Foundation. CSH acknowledges additional support by the Deutsche Forschungsgemeinschaft (DFG, German Research Foundation) under Germany’s Excellence Strategy – EXC 2181/1 - 390900948 (the Heidelberg STRUCTURES Excellence Cluster). SJN is supported by the US National Science Foundation (NSF) through grant AST-2108402, and the Polish National Science Centre through grant UMO-2018/31/N/ST9/03975. JZ is supported by the project “NRW-Cluster for data intensive radio astronomy: Big Bang to Big Data (B3D)“funded through the programme “Profilbildung 2020”, an initiative of the Ministry of Culture and Science of the State of North Rhine-Westphalia. SC acknowledges support from the Italian Ministry of University and Research (\textsc{mur}) through PRIN 2022 'EXSKALIBUR – Euclid-Cross-SKA: Likelihood Inference Building for Universe's Research' and from the European Union -- Next Generation EU.

LOFAR data products were provided by the LOFAR Surveys Key Science project (LSKSP; \url{https://lofar-surveys.org/}) and were derived from observations with the International LOFAR Telescope (ILT). 
LOFAR \citep{LOFAR} is the Low Frequency Array designed and constructed by ASTRON. It has observing, data processing, and data storage facilities in several countries, which are owned by various parties (each with their own funding sources), and which are collectively operated by the ILT foundation under a joint scientific policy. The efforts of the LSKSP have benefited from funding from the European Research Council, NOVA, NWO, CNRS-INSU, the SURF Co-operative, the UK Science and Technology Funding Council and the Jülich Supercomputing Centre. 

This research made use of Astropy, a community developed core Python package for astronomy \citep{Astropy2013} hosted at \url{http://www.astropy.org/}, matplotlib \citep{matplotlib}, NumPy \citep{NumPy}, lmfit \citep{lmfit2016}, TopCat \citep{TOPCAT}, SciPy \citep{SciPy}, h5py \citep{h5py}, TreeCorr \citep{TreeCorr2004} and Python language \citep{python}.
\end{acknowledgements}

%
%

\bibliographystyle{aa}
\bibliography{aa}

\begin{thebibliography}{74}
\expandafter\ifx\csname natexlab\endcsname\relax\def\natexlab#1{#1}\fi

\bibitem[{{Almosallam} {et~al.}(2016{\natexlab{a}}){Almosallam}, {Jarvis}, \&
  {Roberts}}]{Almosallam2016b}
{Almosallam}, I.~A., {Jarvis}, M.~J., \& {Roberts}, S.~J. 2016{\natexlab{a}},
  \mnras, 462, 726

\bibitem[{{Almosallam} {et~al.}(2016{\natexlab{b}}){Almosallam}, {Lindsay},
  {Jarvis}, \& {Roberts}}]{Almosallam2016a}
{Almosallam}, I.~A., {Lindsay}, S.~N., {Jarvis}, M.~J., \& {Roberts}, S.~J.
  2016{\natexlab{b}}, \mnras, 455, 2387

\bibitem[{{Astropy Collaboration} {et~al.}(2013){Astropy Collaboration},
  {Robitaille}, {Tollerud}, {Greenfield}, {Droettboom}, {Bray}, {Aldcroft},
  {Davis}, {Ginsburg}, {Price-Whelan}, {Kerzendorf}, {Conley}, {Crighton},
  {Barbary}, {Muna}, {Ferguson}, {Grollier}, {Parikh}, {Nair}, {Unther},
  {Deil}, {Woillez}, {Conseil}, {Kramer}, {Turner}, {Singer}, {Fox}, {Weaver},
  {Zabalza}, {Edwards}, {Azalee Bostroem}, {Burke}, {Casey}, {Crawford},
  {Dencheva}, {Ely}, {Jenness}, {Labrie}, {Lim}, {Pierfederici}, {Pontzen},
  {Ptak}, {Refsdal}, {Servillat}, \& {Streicher}}]{Astropy2013}
{Astropy Collaboration}, {Robitaille}, T.~P., {Tollerud}, E.~J., {et~al.} 2013,
  \aap, 558, A33

\bibitem[{{Bahcall} \& {Soneira}(1983)}]{Bahcall}
{Bahcall}, N.~A. \& {Soneira}, R.~M. 1983, \apj, 270, 20

\bibitem[{{Becker} {et~al.}(1995){Becker}, {White}, \& {Helfand}}]{Becker95}
{Becker}, R.~H., {White}, R.~L., \& {Helfand}, D.~J. 1995, \apj, 450, 559

\bibitem[{{Ben{\'\i}tez}(2000)}]{Benitez20}
{Ben{\'\i}tez}, N. 2000, \apj, 536, 571

\bibitem[{Best {et~al.}(2023)Best, Kondapally, Williams, Cochrane, Duncan,
  Hale, Haskell, Małek, McCheyne, Smith, Wang, Botteon, Bonato, Bondi, Rivera,
  Gao, Gürkan, Hardcastle, Jarvis, Mingo, Miraghaei, Morabito, Nisbet,
  Prandoni, Röttgering, Sabater, Shimwell, Tasse, \& van Weeren}]{Best_2023}
Best, P.~N., Kondapally, R., Williams, W.~L., {et~al.} 2023, Monthly Notices of
  the Royal Astronomical Society, 523, 1729–1755

\bibitem[{{Blake} {et~al.}(2004){Blake}, {Mauch}, \& {Sadler}}]{Blake2004}
{Blake}, C., {Mauch}, T., \& {Sadler}, E.~M. 2004, \mnras, 347, 787

\bibitem[{{Blake} \& {Wall}(2002)}]{BlakeWall2002}
{Blake}, C. \& {Wall}, J. 2002, \mnras, 337, 993

\bibitem[{Bonaldi {et~al.}(2023)Bonaldi, Hartley, Ronconi, De Zotti, \&
  Bonato}]{10.1093/mnras/stad1913}
Bonaldi, A., Hartley, P., Ronconi, T., De Zotti, G., \& Bonato, M. 2023,
  Monthly Notices of the Royal Astronomical Society, 524, 993

\bibitem[{{Bonvin} \& {Durrer}(2011)}]{BonvinDurrer11}
{Bonvin}, C. \& {Durrer}, R. 2011, \prd, 84, 063505

\bibitem[{{Brammer} {et~al.}(2008){Brammer}, {van Dokkum}, \&
  {Coppi}}]{Brammer2008}
{Brammer}, G.~B., {van Dokkum}, P.~G., \& {Coppi}, P. 2008, \apj, 686, 1503

\bibitem[{Brodwin {et~al.}(2006)Brodwin, Brown, Ashby, Bian, Brand, Dey,
  Eisenhardt, Eisenstein, Gonzalez, Huang, Jannuzi, Kochanek, McKenzie, Murray,
  Pahre, Smith, Soifer, Stanford, Stern, \& Elston}]{Brodwin_2006}
Brodwin, M., Brown, M. J.~I., Ashby, M. L.~N., {et~al.} 2006, \apj, 651, 791

\bibitem[{{Brookes} {et~al.}(2008){Brookes}, {Best}, {Peacock},
  {R{\"o}ttgering}, \& {Dunlop}}]{Brookes_2008}
{Brookes}, M.~H., {Best}, P.~N., {Peacock}, J.~A., {R{\"o}ttgering}, H.~J.~A.,
  \& {Dunlop}, J.~S. 2008, \mnras, 385, 1297

\bibitem[{{Camera} {et~al.}(2012){Camera}, {Santos}, {Bacon}, {Jarvis},
  {McAlpine}, {Norris}, {Raccanelli}, \& {R{\"o}ttgering}}]{Camera2012}
{Camera}, S., {Santos}, M.~G., {Bacon}, D.~J., {et~al.} 2012, \mnras, 427, 2079

\bibitem[{{Challinor} \& {Lewis}(2011)}]{ChallinorLewis11}
{Challinor}, A. \& {Lewis}, A. 2011, \prd, 84, 043516

\bibitem[{{Chen} \& {Schwarz}(2016)}]{ChenSchwarz16}
{Chen}, S. \& {Schwarz}, D.~J. 2016, \aap, 591, A135

\bibitem[{Collete \& {Contributors}(2014)}]{h5py}
Collete, A. \& {Contributors}. 2014

\bibitem[{{Condon} {et~al.}(1998){Condon}, {Cotton}, {Greisen}, {Yin},
  {Perley}, {Taylor}, \& {Broderick}}]{Condon98}
{Condon}, J.~J., {Cotton}, W.~D., {Greisen}, E.~W., {et~al.} 1998, \aj, 115,
  1693

\bibitem[{{Cress} {et~al.}(1996){Cress}, {Helfand}, {Becker}, {Gregg}, \&
  {White}}]{Cress1996}
{Cress}, C.~M., {Helfand}, D.~J., {Becker}, R.~H., {Gregg}, M.~D., \& {White},
  R.~L. 1996, \apj, 473, 7

\bibitem[{{Dahlen} {et~al.}(2013){Dahlen}, {Mobasher}, {Faber}, {Ferguson},
  {Barro}, {Finkelstein}, {Finlator}, {Fontana}, {Gruetzbauch}, {Johnson},
  {Pforr}, {Salvato}, {Wiklind}, {Wuyts}, {Acquaviva}, {Dickinson}, {Guo},
  {Huang}, {Huang}, {Newman}, {Bell}, {Conselice}, {Galametz}, {Gawiser},
  {Giavalisco}, {Grogin}, {Hathi}, {Kocevski}, {Koekemoer}, {Koo}, {Lee},
  {McGrath}, {Papovich}, {Peth}, {Ryan}, {Somerville}, {Weiner}, \&
  {Wilson}}]{Dahlen2013}
{Dahlen}, T., {Mobasher}, B., {Faber}, S.~M., {et~al.} 2013, \apj, 775, 93

\bibitem[{{Dolfi} {et~al.}(2019){Dolfi}, {Branchini}, {Bilicki},
  {Balaguera-Antol{\'\i}nez}, {Prandoni}, \& {Pand it}}]{Dolfi2019}
{Dolfi}, A., {Branchini}, E., {Bilicki}, M., {et~al.} 2019, \aap, 623, A148

\bibitem[{{Donoso} {et~al.}(2009){Donoso}, {Best}, \& {Kauffmann}}]{Donoso2009}
{Donoso}, E., {Best}, P.~N., \& {Kauffmann}, G. 2009, \mnras, 392, 617

\bibitem[{Drinkwater \& Schmidt(1996)}]{drinkwater_schmidt_1996}
Drinkwater, M.~J. \& Schmidt, R.~W. 1996, Publications of the Astronomical
  Society of Australia, 13, 127–131

\bibitem[{{Duncan} {et~al.}(2018{\natexlab{a}}){Duncan}, {Brown}, {Williams},
  {Best}, {Buat}, {Burgarella}, {Jarvis}, {Ma{\l}ek}, {Oliver},
  {R{\"o}ttgering}, \& {Smith}}]{Duncan2018a}
{Duncan}, K.~J., {Brown}, M. J.~I., {Williams}, W.~L., {et~al.}
  2018{\natexlab{a}}, \mnras, 473, 2655

\bibitem[{{Duncan} {et~al.}(2018{\natexlab{b}}){Duncan}, {Jarvis}, {Brown}, \&
  {R{\"o}ttgering}}]{Duncan2018b}
{Duncan}, K.~J., {Jarvis}, M.~J., {Brown}, M. J.~I., \& {R{\"o}ttgering}, H.
  J.~A. 2018{\natexlab{b}}, \mnras, 477, 5177

\bibitem[{{Duncan} {et~al.}(2021){Duncan}, {Kondapally}, {Brown}, {Bonato},
  {Best}, {R{\"o}ttgering}, {Bondi}, {Bowler}, {Cochrane}, {G{\"u}rkan},
  {Hardcastle}, {Jarvis}, {Kunert-Bajraszewska}, {Leslie}, {Ma{\l}ek},
  {Morabito}, {O'Sullivan}, {Prandoni}, {Sabater}, {Shimwell}, {Smith}, {Wang},
  {Wo{\l}owska}, \& {Tasse}}]{Ken21}
{Duncan}, K.~J., {Kondapally}, R., {Brown}, M.~J.~I., {et~al.} 2021, \aap, 648,
  A4

\bibitem[{{Duncan} {et~al.}(2019){Duncan}, {Sabater}, {R{\"o}ttgering},
  {Jarvis}, {Smith}, {Best}, {Callingham}, {Cochrane}, {Croston}, {Hardcastle},
  {Mingo}, {Morabito}, {Nisbet}, {Prandoni}, {Shimwell}, {Tasse}, {White},
  {Williams}, {Alegre}, {Chy{\.z}y}, {G{\"u}rkan}, {Hoeft}, {Kondapally},
  {Mechev}, {Miley}, {Schwarz}, \& {van Weeren}}]{Duncan2019}
{Duncan}, K.~J., {Sabater}, J., {R{\"o}ttgering}, H.~J.~A., {et~al.} 2019,
  \aap, 622, A3

\bibitem[{{Eisenstein} {et~al.}(2001){Eisenstein}, {Annis}, {Gunn}, {Szalay},
  {Connolly}, {Nichol}, {Bahcall}, {Bernardi}, {Burles}, {Castander},
  {Fukugita}, {Hogg}, {Ivezi{\'c}}, {Knapp}, {Lupton}, {Narayanan}, {Postman},
  {Reichart}, {Richmond}, {Schneider}, {Schlegel}, {Strauss}, {SubbaRao},
  {Tucker}, {Vanden Berk}, {Vogeley}, {Weinberg}, \& {Yanny}}]{Eisenstein2001}
{Eisenstein}, D.~J., {Annis}, J., {Gunn}, J.~E., {et~al.} 2001, \aj, 122, 2267

\bibitem[{{Groth} \& {Peebles}(1977)}]{1977ApJ...217..385G}
{Groth}, E.~J. \& {Peebles}, P.~J.~E. 1977, \apj, 217, 385

\bibitem[{{Hale} {et~al.}(2024){Hale}, {Schwarz}, {Best}, {Nakoneczny},
  {Alonso}, {Bacon}, {B{\"o}hme}, {Bhardwaj}, {Bilicki}, {Camera}, {Heneka},
  {Pashapour-Ahmadabadi}, {Tiwari}, {Zheng}, {Duncan}, {Jarvis}, {Kondapally},
  {Magliocchetti}, {Rottgering}, \& {Shimwell}}]{Hale_2023}
{Hale}, C.~L., {Schwarz}, D.~J., {Best}, P.~N., {et~al.} 2024, \mnras, 527,
  6540

\bibitem[{Hunter(2007)}]{matplotlib}
Hunter, J.~D. 2007, Computing in Science \& Engineering, 9, 90

\bibitem[{{Intema} {et~al.}(2017){Intema}, {Jagannathan}, {Mooley}, \&
  {Frail}}]{TGSS2017}
{Intema}, H.~T., {Jagannathan}, P., {Mooley}, K.~P., \& {Frail}, D.~A. 2017,
  \aap, 598, A78

\bibitem[{{Jarvis} {et~al.}(2004){Jarvis}, {Bernstein}, \&
  {Jain}}]{TreeCorr2004}
{Jarvis}, M., {Bernstein}, G., \& {Jain}, B. 2004, \mnras, 352, 338

\bibitem[{{Jones} {et~al.}(2004){Jones}, {Saunders}, {Colless}, {Read},
  {Parker}, {Watson}, {Campbell}, {Burkey}, {Mauch}, {Moore}, {Hartley},
  {Cass}, {James}, {Russell}, {Fiegert}, {Dawe}, {Huchra}, {Jarrett}, {Lahav},
  {Lucey}, {Mamon}, {Proust}, {Sadler}, \& {Wakamatsu}}]{6dFGS}
{Jones}, D.~H., {Saunders}, W., {Colless}, M., {et~al.} 2004, \mnras, 355, 747

\bibitem[{{Kimball} \& {Ivezi{\'c}}(2008)}]{Kimball2008}
{Kimball}, A.~E. \& {Ivezi{\'c}}, {\v{Z}}. 2008, \aj, 136, 684

\bibitem[{Kondapally {et~al.}(2021)Kondapally, Best, Hardcastle, Nisbet,
  Bonato, Sabater, Duncan, McCheyne, Cochrane, Bowler, Williams, Shimwell,
  Tasse, Croston, Goyal, Jamrozy, Jarvis, Mahatma, Röttgering, Smith,
  Wo{\l}owska, Bondi, Brienza, Brown, Brüggen, Chambers, Garrett, Gürkan,
  Huber, Kunert-Bajraszewska, Magnier, Mingo, Mostert, Nikiel-Wroczy{\'{n}
  }ski, O'Sullivan, Paladino, Ploeckinger, Prandoni, Rosenthal, Schwarz,
  Shulevski, Wagenveld, \& Wang}]{Kondapally2021}
Kondapally, R., Best, P.~N., Hardcastle, M.~J., {et~al.} 2021, \aap, 648, A3

\bibitem[{{Landy} \& {Szalay}(1993)}]{LandySzalay1993}
{Landy}, S.~D. \& {Szalay}, A.~S. 1993, \apj, 412, 64

\bibitem[{{Limber}(1954)}]{Limber53}
{Limber}, D.~N. 1954, \apj, 119, 655

\bibitem[{{Lindsay} {et~al.}(2014){Lindsay}, {Jarvis}, {Santos}, {Brown},
  {Croom}, {Driver}, {Hopkins}, {Liske}, {Loveday}, {Norberg}, \&
  {Robotham}}]{Lindsay2014}
{Lindsay}, S.~N., {Jarvis}, M.~J., {Santos}, M.~G., {et~al.} 2014, \mnras, 440,
  1527

\bibitem[{{Longair}(1978)}]{Longair}
{Longair}, M.~S. 1978, in Large Scale Structures in the Universe, ed. M.~S.
  {Longair} \& J.~{Einasto}, Vol.~79, 305

\bibitem[{{Magliocchetti}(2022)}]{2022A&ARv..30....6M}
{Magliocchetti}, M. 2022, \aapr, 30, 6

\bibitem[{{Malz} \& {Hogg}(2022)}]{MalzHogg22}
{Malz}, A.~I. \& {Hogg}, D.~W. 2022, \apj, 928, 127

\bibitem[{{Mauch} \& {Sadler}(2007)}]{MauchSadler2007}
{Mauch}, T. \& {Sadler}, E.~M. 2007, \mnras, 375, 931

\bibitem[{{Mazumder} {et~al.}(2022){Mazumder}, {Chakraborty}, \&
  {Datta}}]{Mazumder2022}
{Mazumder}, A., {Chakraborty}, A., \& {Datta}, A. 2022, \mnras, 517, 3407

\bibitem[{{Miley} \& De~Breuck(2008)}]{Miley}
{Miley}, G. \& De~Breuck, C. 2008, \mbox{Astronomy and Astrophysics} Review,
  15, 67

\bibitem[{{Nakoneczny} {et~al.}(2023){Nakoneczny}, {Alonso}, {Bilicki},
  {Schwarz}, {Hale}, {Pollo}, {Heneka}, {Tiwari}, {Zheng}, {Br{\"u}ggen},
  {Jarvis}, \& {Shimwell}}]{Nakoneczny_2023}
{Nakoneczny}, S.~J., {Alonso}, D., {Bilicki}, M., {et~al.} 2023, arXiv
  e-prints, arXiv:2310.07642

\bibitem[{{Newville} {et~al.}(2016){Newville}, {Stensitzki}, {Allen}, {Rawlik},
  {Ingargiola}, \& {Nelson}}]{lmfit2016}
{Newville}, M., {Stensitzki}, T., {Allen}, D.~B., {et~al.} 2016
  [\eprint[ascl]{1606.014}]

\bibitem[{{Nusser} \& {Tiwari}(2015)}]{NusserTiwari2015}
{Nusser}, A. \& {Tiwari}, P. 2015, \apj, 812, 85

\bibitem[{{Overzier} {et~al.}(2003){Overzier}, {R{\"o}ttgering}, {Rengelink},
  \& {Wilman}}]{Overzier2003}
{Overzier}, R.~A., {R{\"o}ttgering}, H.~J.~A., {Rengelink}, R.~B., \& {Wilman},
  R.~J. 2003, \aap, 405, 53

\bibitem[{{P{\^a}ris} {et~al.}(2018){P{\^a}ris}, {Petitjean}, {Aubourg},
  {Myers}, {Streblyanska}, {Lyke}, {Anderson}, {Armengaud}, {Bautista},
  {Blanton}, {Blomqvist}, {Brinkmann}, {Brownstein}, {Brandt}, {Burtin},
  {Dawson}, {de la Torre}, {Georgakakis}, {Gil-Mar{\'\i}n}, {Green}, {Hall},
  {Kneib}, {LaMassa}, {Le Goff}, {MacLeod}, {Mariappan}, {McGreer}, {Merloni},
  {Noterdaeme}, {Palanque-Delabrouille}, {Percival}, {Ross}, {Rossi},
  {Schneider}, {Seo}, {Tojeiro}, {Weaver}, {Weijmans}, {Y{\`e}che}, {Zarrouk},
  \& {Zhao}}]{SDSSDR14}
{P{\^a}ris}, I., {Petitjean}, P., {Aubourg}, {\'E}., {et~al.} 2018, \aap, 613,
  A51

\bibitem[{{Peacock} \& {Nicholson}(1991)}]{1991MNRAS.253..307P}
{Peacock}, J.~A. \& {Nicholson}, D. 1991, \mnras, 253, 307

\bibitem[{{Peebles}(1980)}]{Peebles1980}
{Peebles}, P.~J.~E. 1980, {The large-scale structure of the universe}
  (Princeton University Press)

\bibitem[{{Planck Collaboration} {et~al.}(2020){Planck Collaboration},
  {Aghanim}, {Akrami}, {Ashdown}, {Aumont}, {Baccigalupi}, {Ballardini},
  {Banday}, {Barreiro}, {Bartolo}, {Basak}, {Battye}, {Benabed}, {Bernard},
  {Bersanelli}, {Bielewicz}, {Bock}, {Bond}, {Borrill}, {Bouchet}, {Boulanger},
  {Bucher}, {Burigana}, {Butler}, {Calabrese}, {Cardoso}, {Carron},
  {Challinor}, {Chiang}, {Chluba}, {Colombo}, {Combet}, {Contreras}, {Crill},
  {Cuttaia}, {de Bernardis}, {de Zotti}, {Delabrouille}, {Delouis}, {Di
  Valentino}, {Diego}, {Dor{\'e}}, {Douspis}, {Ducout}, {Dupac}, {Dusini},
  {Efstathiou}, {Elsner}, {En{\ss}lin}, {Eriksen}, {Fantaye}, {Farhang},
  {Fergusson}, {Fernandez-Cobos}, {Finelli}, {Forastieri}, {Frailis},
  {Fraisse}, {Franceschi}, {Frolov}, {Galeotta}, {Galli}, {Ganga},
  {G{\'e}nova-Santos}, {Gerbino}, {Ghosh}, {Gonz{\'a}lez-Nuevo}, {G{\'o}rski},
  {Gratton}, {Gruppuso}, {Gudmundsson}, {Hamann}, {Handley}, {Hansen},
  {Herranz}, {Hildebrandt}, {Hivon}, {Huang}, {Jaffe}, {Jones}, {Karakci},
  {Keih{\"a}nen}, {Keskitalo}, {Kiiveri}, {Kim}, {Kisner}, {Knox},
  {Krachmalnicoff}, {Kunz}, {Kurki-Suonio}, {Lagache}, {Lamarre}, {Lasenby},
  {Lattanzi}, {Lawrence}, {Le Jeune}, {Lemos}, {Lesgourgues}, {Levrier},
  {Lewis}, {Liguori}, {Lilje}, {Lilley}, {Lindholm}, {L{\'o}pez-Caniego},
  {Lubin}, {Ma}, {Mac{\'\i}as-P{\'e}rez}, {Maggio}, {Maino}, {Mandolesi},
  {Mangilli}, {Marcos-Caballero}, {Maris}, {Martin}, {Martinelli},
  {Mart{\'\i}nez-Gonz{\'a}lez}, {Matarrese}, {Mauri}, {McEwen}, {Meinhold},
  {Melchiorri}, {Mennella}, {Migliaccio}, {Millea}, {Mitra},
  {Miville-Desch{\^e}nes}, {Molinari}, {Montier}, {Morgante}, {Moss}, {Natoli},
  {N{\o}rgaard-Nielsen}, {Pagano}, {Paoletti}, {Partridge}, {Patanchon},
  {Peiris}, {Perrotta}, {Pettorino}, {Piacentini}, {Polastri}, {Polenta},
  {Puget}, {Rachen}, {Reinecke}, {Remazeilles}, {Renzi}, {Rocha}, {Rosset},
  {Roudier}, {Rubi{\~n}o-Mart{\'\i}n}, {Ruiz-Granados}, {Salvati}, {Sandri},
  {Savelainen}, {Scott}, {Shellard}, {Sirignano}, {Sirri}, {Spencer},
  {Sunyaev}, {Suur-Uski}, {Tauber}, {Tavagnacco}, {Tenti}, {Toffolatti},
  {Tomasi}, {Trombetti}, {Valenziano}, {Valiviita}, {Van Tent}, {Vibert},
  {Vielva}, {Villa}, {Vittorio}, {Wandelt}, {Wehus}, {White}, {White},
  {Zacchei}, \& {Zonca}}]{2020A&A...641A...6P}
{Planck Collaboration}, {Aghanim}, N., {Akrami}, Y., {et~al.} 2020, \aap, 641,
  A6

\bibitem[{{Postman} {et~al.}(1992){Postman}, {Huchra}, \& {Geller}}]{Postman}
{Postman}, M., {Huchra}, J.~P., \& {Geller}, M.~J. 1992, \apj, 384, 404

\bibitem[{{Rana} \& {Bagla}(2019{\natexlab{a}})}]{RanaBagla2019}
{Rana}, S. \& {Bagla}, J.~S. 2019{\natexlab{a}}, \mnras, 485, 5891

\bibitem[{{Rana} \& {Bagla}(2019{\natexlab{b}})}]{RanaBaglaErratum}
{Rana}, S. \& {Bagla}, J.~S. 2019{\natexlab{b}}, \mnras, 487, 1821

\bibitem[{{Sabater} {et~al.}(2021){Sabater}, {Best, P. N.}, {Tasse, C.},
  {Hardcastle, M. J.}, {Shimwell, T. W.}, {Nisbet, D.}, {Jelic, V.},
  {Callingham, J. R.}, {R\"ottgering, H. J. A.}, {Bonato, M.}, {Bondi, M.},
  {Ciardi, B.}, {Cochrane, R. K.}, {Jarvis, M. J.}, {Kondapally, R.},
  {Koopmans, L. V. E.}, {O\'{}Sullivan, S. P.}, {Prandoni, I.}, {Schwarz, D.
  J.}, {Smith, D. J. B.}, {Wang, L.}, {Williams, W. L.}, \& {Zaroubi,
  S.}}]{Sabater2021}
{Sabater}, J., {Best, P. N.}, {Tasse, C.}, {et~al.} 2021, A\&A, 648, A2

\bibitem[{{Scoville} {et~al.}(2007){Scoville}, {Aussel}, {Brusa}, {Capak},
  {Carollo}, {Elvis}, {Giavalisco}, {Guzzo}, {Hasinger}, {Impey}, {Kneib},
  {LeFevre}, {Lilly}, {Mobasher}, {Renzini}, {Rich}, {Sanders}, {Schinnerer},
  {Schminovich}, {Shopbell}, {Taniguchi}, \& {Tyson}}]{2007ApJS..172....1S}
{Scoville}, N., {Aussel}, H., {Brusa}, M., {et~al.} 2007, \apjs, 172, 1

\bibitem[{{Shimwell} {et~al.}(2022){Shimwell}, {Hardcastle}, {Tasse}, {Best},
  {R{\"o}ttgering}, {Williams}, {Botteon}, {Drabent}, {Mechev}, {Shulevski},
  {van Weeren}, {Bester}, {Br{\"u}ggen}, {Brunetti}, {Callingham}, {Chy{\.z}y},
  {Conway}, {Dijkema}, {Duncan}, {de Gasperin}, {Hale}, {Haverkorn}, {Hugo},
  {Jackson}, {Mevius}, {Miley}, {Morabito}, {Morganti}, {Offringa}, {Oonk},
  {Rafferty}, {Sabater}, {Smith}, {Schwarz}, {Smirnov}, {O'Sullivan},
  {Vedantham}, {White}, {Albert}, {Alegre}, {Asabere}, {Bacon}, {Bonafede},
  {Bonnassieux}, {Brienza}, {Bilicki}, {Bonato}, {Calistro Rivera}, {Cassano},
  {Cochrane}, {Croston}, {Cuciti}, {Dallacasa}, {Danezi}, {Dettmar}, {Di
  Gennaro}, {Edler}, {En{\ss}lin}, {Emig}, {Franzen}, {Garc{\'\i}a-Vergara},
  {Grange}, {G{\"u}rkan}, {Hajduk}, {Heald}, {Heesen}, {Hoang}, {Hoeft},
  {Horellou}, {Iacobelli}, {Jamrozy}, {Jeli{\'c}}, {Kondapally}, {Kukreti},
  {Kunert-Bajraszewska}, {Magliocchetti}, {Mahatma}, {Ma{\l}ek}, {Mandal},
  {Massaro}, {Meyer-Zhao}, {Mingo}, {Mostert}, {Nair}, {Nakoneczny},
  {Nikiel-Wroczy{\'n}ski}, {Orr{\'u}}, {Pajdosz-{\'S}mierciak}, {Pasini},
  {Prandoni}, {van Piggelen}, {Rajpurohit}, {Retana-Montenegro}, {Riseley},
  {Rowlinson}, {Saxena}, {Schrijvers}, {Sweijen}, {Siewert}, {Timmerman},
  {Vaccari}, {Vink}, {West}, {Wo{\l}owska}, {Zhang}, \& {Zheng}}]{Shimwell2022}
{Shimwell}, T.~W., {Hardcastle}, M.~J., {Tasse}, C., {et~al.} 2022, \aap, 659,
  A1

\bibitem[{{Shimwell} {et~al.}(2019){Shimwell}, {Tasse}, {Hardcastle}, {Mechev},
  {Williams}, {Best}, {R{\"o}ttgering}, {Callingham}, {Dijkema}, {de Gasperin},
  {Hoang}, {Hugo}, {Mirmont}, {Oonk}, {Prandoni}, {Rafferty}, {Sabater},
  {Smirnov}, {van Weeren}, {White}, {Atemkeng}, {Bester}, {Bonnassieux},
  {Br{\"u}ggen}, {Brunetti}, {Chy{\.z}y}, {Cochrane}, {Conway}, {Croston},
  {Danezi}, {Duncan}, {Haverkorn}, {Heald}, {Iacobelli}, {Intema}, {Jackson},
  {Jamrozy}, {Jarvis}, {Lakhoo}, {Mevius}, {Miley}, {Morabito}, {Morganti},
  {Nisbet}, {Orr{\'u}}, {Perkins}, {Pizzo}, {Schrijvers}, {Smith}, {Vermeulen},
  {Wise}, {Alegre}, {Bacon}, {van Bemmel}, {Beswick}, {Bonafede}, {Botteon},
  {Bourke}, {Brienza}, {Calistro Rivera}, {Cassano}, {Clarke}, {Conselice},
  {Dettmar}, {Drabent}, {Dumba}, {Emig}, {En{\ss}lin}, {Ferrari}, {Garrett},
  {G{\'e}nova-Santos}, {Goyal}, {G{\"u}rkan}, {Hale}, {Harwood}, {Heesen},
  {Hoeft}, {Horellou}, {Jackson}, {Kokotanekov}, {Kondapally},
  {Kunert-Bajraszewska}, {Mahatma}, {Mahony}, {Mandal}, {McKean}, {Merloni},
  {Mingo}, {Miskolczi}, {Mooney}, {Nikiel-Wroczy{\'n}ski}, {O'Sullivan},
  {Quinn}, {Reich}, {Roskowi{\'n}ski}, {Rowlinson}, {Savini}, {Saxena},
  {Schwarz}, {Shulevski}, {Sridhar}, {Stacey}, {Urquhart}, {van der Wiel},
  {Varenius}, {Webster}, \& {Wilber}}]{Shimwell19}
{Shimwell}, T.~W., {Tasse}, C., {Hardcastle}, M.~J., {et~al.} 2019, \aap, 622,
  A1

\bibitem[{{Siewert} {et~al.}(2020){Siewert}, {Hale, C.}, {Bhardwaj, N.},
  {Biermann, M.}, {Bacon, D. J.}, {Jarvis, M.}, {R\"ottgering, H. J .A.},
  {Schwarz, D. J.}, {Shimwell, T.}, {Best, P. N.}, {Duncan, K. J.},
  {Hardcastle, M. J.}, {Sabater, J.}, {Tasse, C.}, {White, G. J.}, \&
  {Williams, W. L.}}]{siewert}
{Siewert}, T.~M., {Hale, C.}, {Bhardwaj, N.}, {et~al.} 2020, A\&A, 643, A100

\bibitem[{Simon(2007)}]{Simon2007}
Simon, P. 2007, \aap, 473, 711–714

\bibitem[{{Smith} {et~al.}(2016){Smith}, {Best}, {Duncan}, {Hatch}, {Jarvis},
  {R{\"o}ttgering}, {Simpson}, {Stott}, {Cochrane}, {Coppin}, {Dannerbauer},
  {Davis}, {Geach}, {Hale}, {Hardcastle}, {Hatfield}, {Houghton}, {Maddox},
  {McGee}, {Morabito}, {Nisbet}, {Pandey-Pommier}, {Prandoni}, {Saxena},
  {Shimwell}, {Tarr}, {van Bemmel}, {Verma}, {White}, \&
  {Williams}}]{2016sf2a.conf..271S}
{Smith}, D.~J.~B., {Best}, P.~N., {Duncan}, K.~J., {et~al.} 2016, in SF2A-2016:
  Proceedings of the Annual meeting of the French Society of Astronomy and
  Astrophysics, ed. C.~{Reyl{\'e}}, J.~{Richard}, L.~{Cambr{\'e}sy},
  M.~{Deleuil}, E.~{P{\'e}contal}, L.~{Tresse}, \& I.~{Vauglin}, 271--280

\bibitem[{{Smolčić} {et~al.}(2017){Smolčić}, {Delvecchio, I.}, {Zamorani,
  G.}, {Baran, N.}, {Novak, M.}, {Delhaize, J.}, {Schinnerer, E.}, {Berta, S.},
  {Bondi, M.}, {Ciliegi, P.}, {Capak, P.}, {Civano, F.}, {Karim, A.}, {Le
  Fevre, O.}, {Ilbert, O.}, {Laigle, C.}, {Marchesi, S.}, {McCracken, H. J.},
  {Tasca, L.}, {Salvato, M.}, \& {Vardoulaki, E.}}]{refId0}
{Smolčić}, V., {Delvecchio, I.}, {Zamorani, G.}, {et~al.} 2017, A\&A, 602, A2

\bibitem[{{Tasse} {et~al.}(2021){Tasse}, {Shimwell, T.}, {Hardcastle, M. J.},
  {O\'{}Sullivan, S. P.}, {van Weeren, R.}, {Best, P. N.}, {Bester, L.}, {Hugo,
  B.}, {Smirnov, O.}, {Sabater, J.}, {Calistro-Rivera, G.}, {de Gasperin, F.},
  {Morabito, L. K.}, {R\"ottgering, H.}, {Williams, W. L.}, {Bonato, M.},
  {Bondi, M.}, {Botteon, A.}, {Br\"uggen, M.}, {Brunetti, G.}, {Chyzy, K. T.},
  {Garrett, M. A.}, {G\"urkan, G.}, {Jarvis, M. J.}, {Kondapally, R.}, {Mandal,
  S.}, {Prandoni, I.}, {Repetti, A.}, {Retana-Montenegro, E.}, {Schwarz, D.
  J.}, {Shulevski, A.}, \& {Wiaux, Y.}}]{Tasse2021}
{Tasse}, C., {Shimwell, T.}, {Hardcastle, M. J.}, {et~al.} 2021, A\&A, 648, A1

\bibitem[{{Taylor}(2005)}]{TOPCAT}
{Taylor}, M.~B. 2005, Astronomical Society of the Pacific Conference Series,
  Vol. 347, {TOPCAT \& STIL: Starlink Table/VOTable Processing Software}, ed.
  P.~{Shopbell}, M.~{Britton}, \& R.~{Ebert}, 29

\bibitem[{{van Haarlem} {et~al.}(2013){van Haarlem}, Wise, Gunst, Heald,
  McKean, Hessels, {de Bruyn}, Nijboer, Swinbank, Fallows, Brentjens, Nelles,
  Beck, Falcke, Fender, H{\"o}randel, Koopmans, Mann, Miley, R{\"o}ttgering,
  Stappers, Wijers, Zaroubi, {van den Akker}, Alexov, Anderson, Anderson, {van
  Ardenne}, Arts, Asgekar, Avruch, Batejat, B{\"a}hren, Bell, Bell, {van
  Bemmel}, Bennema, Bentum, Bernardi, Best, B{\^i}rzan, Bonafede, Boonstra,
  Braun, Bregman, Breitling, {van de Brink}, Broderick, Broekema, Brouw,
  Br{\"u}ggen, Butcher, {van Cappellen}, Ciardi, Coenen, Conway, Coolen,
  Corstanje, Damstra, Davies, Deller, Dettmar, {van Diepen}, Dijkstra, Donker,
  Doorduin, Dromer, Drost, {van Duin}, Eisl{\"o}ffel, {van Enst}, Ferrari,
  Frieswijk, Gankema, Garrett, {de Gasperin}, Gerbers, {de Geus},
  Grie{\ss}meier, Grit, Gruppen, Hamaker, Hassall, Hoeft, Holties, Horneffer,
  {van der Horst}, {van Houwelingen}, Huijgen, Iacobelli, Intema, Jackson,
  Jelic, {de Jong}, Juette, Kant, Karastergiou, Koers, Kollen, Kondratiev,
  Kooistra, Koopman, Koster, Kuniyoshi, Kramer, Kuper, Lambropoulos, Law, {van
  Leeuwen}, Lemaitre, Loose, Maat, Macario, Markoff, Masters, McFadden,
  {McKay-Bukowski}, Meijering, Meulman, Mevius, Middelberg, Millenaar,
  {Miller-Jones}, Mohan, Mol, Morawietz, Morganti, Mulcahy, Mulder, Munk,
  Nieuwenhuis, {van Nieuwpoort}, Noordam, Norden, Noutsos, Offringa, Olofsson,
  Omar, Orr{\'u}, Overeem, Paas, {Pandey-Pommier}, Pandey, Pizzo, Polatidis,
  Rafferty, Rawlings, Reich, {de Reijer}, Reitsma, Renting, Riemers, Rol,
  Romein, Roosjen, Ruiter, Scaife, {van der Schaaf}, Scheers, Schellart,
  Schoenmakers, Schoonderbeek, Serylak, Shulevski, Sluman, Smirnov, Sobey,
  Spreeuw, Steinmetz, Sterks, Stiepel, Stuurwold, Tagger, Tang, Tasse, Thomas,
  Thoudam, Toribio, {van der Tol}, Usov, {van Veelen}, {van der Veen}, {ter
  Veen}, Verbiest, Vermeulen, Vermaas, Vocks, Vogt, {de Vos}, {van der Wal},
  {van Weeren}, Weggemans, Weltevrede, White, Wijnholds, Wilhelmsson, Wucknitz,
  Yatawatta, Zarka, Zensus, \& {van Zwieten}}]{LOFAR}
{van Haarlem}, M.~P., Wise, M.~W., Gunst, A.~W., {et~al.} 2013, \aap, 556, A2

\bibitem[{van Rossum(1995)}]{python}
van Rossum, G. 1995

\bibitem[{{Virtanen} {et~al.}(2020){Virtanen}, {Gommers}, {Oliphant},
  {Haberland}, {Reddy}, {Cournapeau}, {Burovski}, {Peterson}, {Weckesser},
  {Bright}, {van der Walt}, {Brett}, {Wilson}, {Jarrod Millman}, {Mayorov},
  {Nelson}, {Jones}, {Kern}, {Larson}, {Carey}, {Polat}, {Feng}, {Moore}, {Vand
  erPlas}, {Laxalde}, {Perktold}, {Cimrman}, {Henriksen}, {Quintero}, {Harris},
  {Archibald}, {Ribeiro}, {Pedregosa}, {van Mulbregt}, \&
  {Contributors}}]{SciPy}
{Virtanen}, P., {Gommers}, R., {Oliphant}, T.~E., {et~al.} 2020, Nature
  Methods, 17, 261

\bibitem[{Walt {et~al.}(2011)Walt, Colbert, \& Varoquaux}]{NumPy}
Walt, S. v.~d., Colbert, S.~C., \& Varoquaux, G. 2011, Computing in Science \&
  Engineering, 13, 22

\bibitem[{{Williams} {et~al.}(2019){Williams}, {Hardcastle}, {Best}, {Sabater},
  {Croston}, {Duncan}, {Shimwell}, {R{\"o}ttgering}, {Nisbet}, {G{\"u}rkan},
  {Alegre}, {Cochrane}, {Goyal}, {Hale}, {Jackson}, {Jamrozy}, {Kondapally},
  {Kunert-Bajraszewska}, {Mahatma}, {Mingo}, {Morabito}, {Prandoni},
  {Roskowinski}, {Shulevski}, {Smith}, {Tasse}, {Urquhart}, {Webster}, {White},
  {Beswick}, {Callingham}, {Chy{\.z}y}, {de Gasperin}, {Harwood}, {Hoeft},
  {Iacobelli}, {McKean}, {Mechev}, {Miley}, {Schwarz}, \& {van
  Weeren}}]{Williams2019}
{Williams}, W.~L., {Hardcastle}, M.~J., {Best}, P.~N., {et~al.} 2019, \aap,
  622, A2

\bibitem[{{Wittman} {et~al.}(2016){Wittman}, {Bhaskar}, \&
  {Tobin}}]{Wittman2016}
{Wittman}, D., {Bhaskar}, R., \& {Tobin}, R. 2016, \mnras, 457, 4005

\bibitem[{{Yoo}(2010)}]{Yoo10}
{Yoo}, J. 2010, \prd, 82, 083508

\end{thebibliography}

\end{document}